\documentclass[12pt]{article}
\usepackage[margin = 1in]{geometry}
\usepackage[utf8]{inputenc}
\usepackage{graphicx} % Extended graphics package.

\usepackage{amsmath} % American Mathematics Society standards
\usepackage{amsxtra} % Additional math symbols
\usepackage{amssymb} % Additional math symbols
\usepackage{amsthm} % Additional math symbols
\usepackage{amsfonts}
\usepackage{latexsym} % Additional math symbols
\usepackage{tikz}
\usetikzlibrary {shapes.multipart}
\usetikzlibrary{matrix}
\usepackage{caption}
\usepackage{subcaption}
\usepackage{comment}
\usepackage[authoryear]{natbib}
\usepackage{setspace}

\newcommand{\nn}{\nonumber \\}

\DeclareMathOperator*{\argmax}{arg\,max}
\DeclareMathOperator*{\argmin}{arg\,min}

\begin{document}

\title{\bf Demystifying Statistical Matching Algorithms for Big Data}
  \author{
  	Sanjeewani Weerasingha\\
    Department of Statistics, The Ohio State University\\ 
    and \\
    Michael J. Higgins \\
    Department of Statistics, Kansas State University}
\maketitle

\begin{abstract}

Statistical matching is an effective method for estimating causal effects in which treated units are paired with control units with ``similar'' values of confounding covariates prior to performing estimation.
In this way, matching helps isolate the effect of treatment on response from effects due to the confounding covariates.
While there are a large number of software packages to perform statistical matching, the algorithms and techniques used to solve statistical matching problems---especially matching without replacement---are not widely understood.
In this paper, we describe in detail commonly-used algorithms and techniques for solving statistical matching problems.
We focus in particular on the efficiency of these algorithms as the number of observations grow large.
We advocate for the further development of statistical matching methods that impose and exploit ``sparsity''---by greatly restricting the available matches for a given treated unit---as this may be critical to ensure scalability of matching methods as data sizes grow large.
\end{abstract}

%MIKE: NEED TO CITE THE FIGURES

% %MIKE: Adding Big Data
% \title{Demystifying Statistical Matching Algorithms for Big Data}
% \author{Sanjeewani Weerasingha, Michael Higgins
% }
% \date{May 2022}

% \begin{document}

% \maketitle
% % +--------------------------------------------------------------------+
% % | Sample Chapter 2
% % +--------------------------------------------------------------------+

% \cleardoublepage

% +--------------------------------------------------------------------+
% | Replace "This is Chapter 2" below with the title of your chapter.
% | LaTeX will automatically number the chapters.                      
% +--------------------------------------------------------------------+

%MIKE: Format into Journal of Causal Inference journal format.
%MIKE: Write as \mathbb {I}_T instead of \mathbb{I_T}, same for C' and C.  I don't want the T to be ]mathbb'd.
%MIKE: Make sure, if we remove treated units, to describe changes in the estimand.

%MIKE: Need to make sure we're consistent about lowercase and uppercase N.

\section{Introduction}

\label{makereference2}

Consider an observational study where each unit is given exactly one of two treatment conditions: treatment or control.
%MIKE: CITE of something after sentence
When confounding variables---those that are correlated with both treatment and response---are present, failure to account for this confounding may lead to significant bias in treatment effect estimates~\citep{rosenbaum2010design}.
%MIKE: Need a cite here.
For instance, in a study assessing the effect of smoking and heart disease, confounders include having poor diet and exercise habits as both variables are correlated with an increased incidence in heart disease and a higher likelihood of smoking. 

Statistical matching is a technique designed to isolate the effect of treatment in the presence of confounders.
In statistical matching, treated units are matched with control units with similar values for confounding covariates. 
Treatment effect estimates can then be obtained by taking, for example, the average of the differences in response between the treated and matched control units.
Statistical matching plays an essential role in conducting research work in many subject areas, such as medicine, economics, and political science, since experiments are not always practical or ethical to conduct. 

With advances in computing, the volume of observational data has increased dramatically.
For example, Electronic Health Records (EHR) collect valuable clinical information that researchers can use to guide patient care. 
EHRs include information on patient demographics, progress notes, problem lists, medications, vital signs, past medical history, etc.~\citep{gliklich2019tools}. 
%While these databases may be helpful in the context of patient registries, their applicability and limitations vary widely depending on the size and scope of the database and the research question(s) of interest.
%MIKE: A couple examples here, including electronic medical records.
With this surge in available data, there is a significant need for  matching methods that can be applied under big data settings.

We aim to provide a detailed description of the available techniques and tools for statistical matching, thereby adding some clarity to the black box of statistical matching and possibly enlightening the path towards future advances.
We have particular focus on issues of the scalability of matching algorithms---the ability to successfully apply matching algorithms as the number of units under study becomes large.

%MIKE: This paragraph will be altered over time.  Also use ~\ref{} to refer to sections.
%Section 2.1 review the background materials and notation about statistical matching. Matching problems are well-studied optimization problems in the Operation Research area. In particular, the bipartite matching or statistical matching can be considered as a version of a linear assignment problem in the Operation Research area. Hence,Section 2.2 discusses materials on related materials from the Optimization area for statistical matching. And Section 2.3 discusses how to model a bipartite matching problem as a network flow problem in an optimization framework. Section 2.4 will explore why matching on a sparse graph is important and the existing approaches to solve minimum cost maximum matching on a sparse graph. Finally, Section 2.6 will review helpful algorithms

This chapter is organized as follows. Section 2.1 review the background materials and notation about statistical matching. Matching problems are well-studied optimization problems in the Operation Research area. In particular, the bipartite matching or statistical matching can be considered a version of a linear assignment problem in the Operation Research area. Hence, Section 2.2 discusses materials on related materials from the Optimization area for statistical matching. And Section 2.3 discusses how to model a bipartite matching problem as a network flow problem in an optimization framework. Section 2.4 will explore why matching on a sparse graph is important and the existing approaches to solve minimum cost maximum matching on a sparse graph. Finally, Section 2.6 will demystify the matching algorithms. Moreover, Section 2.6 will discuss helpful materials from the optimization theory area to understand algorithms used in statistical matching.

\section{Problem Setup for Statistical Matching}

%MIKE: I am going to try to write this without referring to \mathbb terms
Consider an observational study on $N$ units, numbered 1 through $N$.  
For each unit $i$, we observe a response $y_i$, a treatment status $T_i \in \{0,1\}$---where $T_i = 1$ denotes that $i$ is given treatment and $T_i = 0$ denotes that $i$ is given control---and a $p$-dimensional vector of confounding covariates $\mathbf x_i = (x_{i1}, x_{i2},\ldots,x_{ip})$.
%MIKE: I am removing the c' in the notation, making it N_T
Let $N_T$ denote the number of treated units, numbered $1$ through $N_T$, and let $N_{C}$ denote the number of control units, numbered $1$ through $N_{C}$.  
For ease of exposition, we assume $N_T \leq N_{C}$.
%Let $\mathbb{I}_T$ and $\mathbb{I}_{C'}$ denote the set of treated and control units respectively.  

Between each treated unit $i$ and control unit $j$, 
%pair of treated and control units $i \in \mathbb {I}_T$, $j \in \mathbb I_{C'}$, 
a dissimilarity measure $w_{ij}$ may be computed on the confounding covariates $\mathbf x$, where smaller values of $w_{ij}$ indicate that $i$ and $j$ have more similar values of confounding covariates.
%MIKE: CITE Imbens and Rubin?
Common choices of $w_{ij}$ include the standardized Euclidean and Mahalanobis distances and the absolute difference in estimated propensity scores~\citep{imbens2015causal}.
Intuitively, matching aims to find, for each treated unit $i$, one or more control units $j$ that have ``small dissimilarity.''
%\in \mathbb I_{T}$, one or more matches $j \in \mathbb I_{C'}$ that have ``small dissimilarity.'' 
%Intuitively, matching aims to find, for each treated unit $i \in \mathbb I_{T}$, one or more matches $j \in \mathbb I_{C'}$ that have ``small dissimilarity.''  

As~\citet{rosenbaum1989optimal} astutely noted, in considering the problem of statistical matching, there is a large body of literature---historically in the field of operations research---on similar types of matching problems from which to draw inspiration. 
%MIKE: Need cites
In deliberately vague terms, these problems often start by assuming a mathematical graph and aim to select connected pairs of units within the graph in an ``optimal'' way. 
Hence, to make statistical matching problems more precise and to help draw connections between statistical matching and matching problems in the operations research literature, we describe these problems in terms graph theory.
%MIKE: Add cites to these types of matching.  Ben Hansen for Full matching, Savje et. al. for Generalized Full matching.
%MIKE: This is incorrect cite placement.   Put them in the examples.
For simplicity, we focus on the 1:1 matching case, where each treated unit is allowed to be matched to, at most, one control unit~\citep{savje2021generalized}, and extend our approach to more complicated matching schemes (e.g. 1:$k$ matching, full matching, generalized full matching, cardinality matching) when appropriate~\citep{hansen2007optmatch}.

For statistical matching, units under study are represented as a graph $G=(V,E)$; each node $i$ in the node set $V$ represents a unit under study (hence, $|V| = N$), and edges $ij$ in the edge set $E$ are drawn between two nodes if their corresponding units are allowed to be matched with each other.  
Each edge $ij$ has a non-negative cost $w_{ij}$ equal to the dissimilarity between the corresponding units $i$ and $j$.
The resulting graph is a \textit{bipartite} graph; the node set $V$ can be partitioned into two groups $V_T$ and $V_{C}$---those nodes that correspond to treated and control units respectively---and edges are only allowed to connect a node from $V_T$ to one in $V_{C}$ (e.g.~you cannot match a treated unit to another treated unit nor a control unit to another control unit).
When initializing a matching problem, it is common to make minimal assumptions on which units can be matched to each other, thereby allowing the matching algorithm to completely determine which matches are appropriate.
In terms of the graph $G$, this corresponds to the assumption that edges $ij \in E$ exist between each pair of units $i \in V_T, j \in V_C$---that is, $G$ is a \textit{complete} bipartite graph. 
For ease of exposition, we may refer to nodes as units and edge costs as dissimilarities throughout this paper.
%MIKE: A cool picture going from an observational study to a graph would be cool here.

A 1:1 statistical matching is a subset of edges $M \subset E$ such that each treated node $i \in V_T$ is \textit{incident} to, at most, one edge $ij \in M$---that is, each node $i\in V_T$ is the endpoint of at most one edge in $M$.  
If $ij \in M$, then the control unit $j$ is matched to the treated unit $i$ before performing analyses.
Since the power of a study is most affected by the number of observations included in the study, often we aim to select a matching $M$ with a large cardinality $|M|$; often, we require the cardinality to be maximized.
%maximized.  $|M|$ that it makes sense to have the cardinality of the matching $|M|$ to be large---or often maximized---in the setup of the matching problem.

%MIKE: For Mike: This will be a paragraph devoted to the objective function.
For a given dataset, there may be a large number of candidate matchings $\mathbb M$ from which to choose.  
%multiple statistical matchings may be considered by a researcher.  
Hence, many statistical matching algorithms aim to select a matching $M^\dagger \in \mathbb M$ that is optimal with respect to some \textit{objective function}.
The commonly used objective function in statistical matching is to minimize total dissimilarity, or cost, between treatment and control pairs  in the matched sample; matching algorithms aim to minimize the total cost
%The most common objective function is the total dissimilarity of edges included in the matching; matching algorithms aim to minimize the total dissimilarity
\begin{equation}
    \label{eq:totalcost}
    M^\dagger = \argmin_{M \in \mathbb M} \sum_{ij \in M} w_{ij}.
\end{equation}
%MIKE: Sanjeewani, please check that this sentence reads OK.  ALso, can you think of other objectives?  I am thinking sample moments may be useful, but I can't think of a lot of methods that use that as the objective rather than a constraint (e.g. CBPS).
Other objectives used in matching include: minimizing the maximum cost within a match~\citep{savje2021generalized}; minimizing the maximum $p$-value for tests for the null hypothesis that covariate distributions between treated and matched control groups are equivalent~\citep{diamond2013genetic}; and maximizing the number of matched pairs subject to constraints on the difference in sample moments between treated and matched control groups~\citep{zubizarreta2012using, zubizarreta2014matching}. 
%and  Given a set of eligible matchings $\mathbb M$, many statistical matching problems aims to find

Matching can be performed \textit{with replacement}---multiple treated units are allowed to be matched to the same control units---or \textit{without replacement}.  
The statistical problem of finding a matching without replacement and the operations research problem of finding a bipartite matching are equivalent.
Hence, significant progress on the statistical matching problem can be made by importing well-studied ideas from the optimization literature.

%Sanjeewani: 246- couple of, 242 : statistical problem, 248: full details of with replacement or without replacement?
Before we discuss optimal methods for performing matching without replacement, we take a couple of brief detours.
First, we describe greedy matching, which is computationally efficient but can suffer from arbitrarily poor performance when matching without replacement.
Then, we discuss in full detail the problem of matching with replacement.
Statistical matching with replacement is a well-understood problem---greedy algorithms can often obtain an optimal matching straightforwardly and efficiently---but it may be inappropriate to use under certain settings.

\section{Greedy Matching \label{sec:GreedMatch}}

%MIKE: Need a paragraph here to describe greedy matching, including pitfalls.  Natural papers to cite include that dehejia and wahba paper and Rosenbaums optimal matching paper.
Greedy algorithms provide a simple and intuitive solution for statistical matching problems. 
Greedy matching algorithms match each treated unit with the eligible control unit that is most similar (with respect to the dissimilarity measure $w$).  
Simple implementations of greedy algorithms can terminate quickly.
Specifically, for each treated unit, the problem of finding the most similar control unit requires $O(N)$ time, and this problem is solved a maximum of $O(N)$ times, leading to a worst-case total runtime of $O(N^2)$~\citep{cormen2022introduction}, outside of the cost of computing the dissimilarities $w$. 
Thus, greedy matching is computationally inexpensive enough for most studies using observational data.
%provided that dissimilarities $w$ can be computed efficiently.
%this process is performed a maximum of $O(N)$ times,
%, and thus, these types of algorithms are often feasible for large datasets.  
%for each treated unit, finding the most similar control unit takes $O(N)$ time, and this process is performed a maximum of $O(N)$ times, and sofinding a maximum value out of $n$ 

However, when matching without replacement, greedy matching may have significant drawbacks.
When the selected dissimilarity measure does not satisfy the triangle inequality (\textit{i.e.}~for any three units $i, j, k$, $w_{ij} + w_{jk} \leq w_{ik}$), the total cost of a 1:1 greedy matching can be infinitely bigger than that for an optimal matching~\citep{rosenbaum1989optimal}.  
Even when the dissimilarity measure satisfies the triangle inequality, the difference in total cost between 1:1 greedy matching and optimal matching may worsen as data sizes get large---to be exact, the difference may be as large as $O(N^{\log_2(3/2)}) \approx O(N^{0.58})$~\citep{reingold1981greedy,agarwal2014approximation}.
Additionally, a greedy matching may have a smaller cardinality than an optimal matching~\citep{rosenbaum1989optimal}, and the matching quality may depend on the order in which treatment units are selected for matching~\citep{dehejia2002propensity}.

A 1:1 greedy matching algorithm proceeds as follows.
\begin{enumerate}
    \item \label{Step0GM} \textbf{(Initialize)} Set the greedy matching $M = \emptyset$.
    \item \label{Step1GM} \textbf{(Select treated node)} Select unit $i \in V_T$ (for example, at random).
    \item \label{Step2GM} \textbf{(Find control match)} Of all eligible control units $j$, find the unit $j^\dagger \in V_C$ that is the most similar to $i$:
    \begin{equation}
        j^\dagger = \argmin_{j: ij \in E} w_{ij}.
    \end{equation}
    Match $i$ to $j^\dagger$: Set $M \xleftarrow{set} M \cup ij^\dagger$.  If no matches are possible, skip to Step~\ref{Step3GM}.
    \item \label{Step3GM} \textbf{(Remove matches)} 
    Set $V_T \xleftarrow{set} V_T \setminus \{i\}$.  
    If matching without replacement, and if a control match $j^\dagger$ was found in Step~\ref{Step2GM}, set $V_C \xleftarrow{set} V_C \setminus \{j^\dagger\}$ and $E \xleftarrow{set} E \setminus \{ij^\dagger : i \in V_C\}$.
    \item \textbf{(Terminate)} 
    If $V_T = \emptyset$, stop.  The matching $M$ is a greedy matching.
    Otherwise, return to Step~\ref{Step1GM}.
\end{enumerate}

A greedy 1:$k$ matching is performed by choosing the $k$ most similar units to the treated unit $i$ in Step~\ref{Step2GM} of the algorithm.
The performance of greedy matching without replacement highly depends on the order in which treated units are selected for matching in Step~\ref{Step1GM}.  
%MIKE: This needs a cite.  I go with Ji and Mitchell's K-Way Equipartition paper (which i think actually needs to be the thesis) for this type of algorithm.
Improved methods for choosing treated nodes---for example, finding the edge $ij \in E$ with the largest cost $w_{ij}$, and choosing the treated unit $i$ incident to this edge---often come with an increased computational cost.

\section{Statistical Matching with Replacement}
%MIKE: Need to make this more concise.  Use fewer sentences but contain the same information.  First paragraph is benefits, second paragraph is drawbacks.

Matching with replacement permits different treated units to be matched to the same control unit.
%MIKE: Do I need proofs of these claims?
%Sanjeewani: as it is optimal?
The biggest advantage of matching with replacement is computational cost.
Greedy matching is almost always used to perform matching with replacement as it is optimal for a number of commonly used objective functions---including the total cost and the maximum cost---under this setting.   

There may be additional instances where, in practice, matching with replacement outperforms without replacement.
For example, matching with replacement may perform better in practice when the distribution of confounding covariates between treated and control groups have little overlap~\citep{dehejia2002propensity}.
It can also be used to estimate the average treatment effect for the treated (ATT) when the number of treated units is greater than the number of control units---matching without replacement would necessarily leave some treated units unmatched, thereby changing the estimand. 
Monte Carlo simulations have suggested that matching with replacement can provide reliable treatment effect estimates if control units are reused for matches infrequently, and suggest that covariate distributions between treated and control groups are more similar for 1:$k$ matching with replacement than without replacement, $k > 1$~\citep{bottigliengo2021oversampling}.

% Additionally, se1:$k$ matching, $k > 1$
% Additionally, 
% A combination of matching with replacement and oversampling ($1:k$ matching where $k>1$) can increase the precision of causal effect estimation. Additionally, a higher level of oversampling($1:K$ where $K>1$) results in a lower covariance imbalance in the matched sample~\citep{bottigliengo2021oversampling}. Since matching with replacement discards fewer treatment units than matching without replacement, it can reduce the bias due to incomplete matching. Thus, the treatment effect estimation can be generalized to the observed data population without bounding it to the matched treatment units.

However, there may also be some drawbacks with matching with replacement.
First and foremost, there is no way to easily control how many times one control unit is used in a match.  
For a given study, it may be possible that many treatment units are matched to a single control unit.
In this case, the response of the control unit will disproportionately influence the estimate of the treatment effect, thereby inflating the standard error of the matching estimator. 
Moreover, simulation results suggest that 1:1 matching without replacement usually yields a smaller difference in sample means between treated and matched control groups than with replacement~\citep{bottigliengo2021oversampling}.
Matching with replacement is rarely used in certain areas of study---for example, in the biomedical sciences---where matching without replacement appears to be more effective~\citep{austin2014use}.

\section{Optimal Statistical Matching Without Replacement}
In 1:1 matching without replacement, each control unit is included in at most one pair in the matched sample. 
Hence, once a control unit is selected for matching, that control unit is no longer eligible for consideration as a potential match for subsequent treatment units.
This substantially increases the difficulty of finding an optimal match, for example, with respect to the total cost objective.  
Thankfully, these types of statistical matching problems are well studied, though most of this work originates from the field of operations research---\citet{rosenbaum1989optimal} first identified the connection between statistical matching and this optimization literature.

The most common statistical matching optimization problem is to  
%problem for optimal 1:1 statistical matching without replacement is to 
find a matching $M^\dagger$ that minimizes the total cost given that it contains as many matched pairs as possible---or more precisely, under the constraint that the cardinality of $M^\dagger$ is maximized.
In the statistical matching literature, these matchings are simply called \textit{optimal matchings}~\citep{rosenbaum1989optimal}.
%MIKE: Start discussing the linear assignment problem here.  Need a cite.
In the optimization literature, this problem is known as the \textit{linear unbalanced assignment problem} (LUAP)~\citep{bijsterbosch2010solving, burkard2012assignment}.

\subsection{The Linear Assignment Problem}
\label{sec:lap}
%Programming problems are concerned with the efficient use or allocation of limited resources to meet desired objectives. 
We begin with a simplification of LUAP---the \textit{linear assignment problem (LAP)}---in which we aim to find an optimal matching $M^\dagger$ when the number of treated units is equal to the number of control units, that is, $N_T = N_C = N/2$.  
In full generality, LAP can be formulated as an \textit{integer linear programming} problem (ILP). However, as we will see, LAP can be solved for small matching problems using a pen and paper.

%Sanjeewani: M or $M^\dagger$ here or what M here refers. 
The ILP formulation of LAP associates each edge $ij \in E$ with a binary variable $z_{ij}$.  
These binary variables \textit{induce} a matching $M$: if $z_{ij}=1$, then the match $ij \in M $, and if $z_{ij}=0$, then $ij \notin M$.
LAP aims to find, across all possible vectors of \textit{variables} $\mathbf z = (z_{ij})_{ij \in E}$, a vector $\textbf{z}^\dagger$ that satisfies
$$
        \mathbf z^\dagger = \argmin_{\mathbf z} \sum_{ij\in E} w_{ij}z_{ij}
$$
under the constraints that
\begin{align}
        \sum_{i \in V_T} z_{ij} = 1&~~~~ \forall ~j \in V_C, \nn
        \sum_{j \in V_C} z_{ij} = 1&~~~~ \forall ~i \in V_T, \label{constraintC}\\
        z_{ij} \in \{0,1\}& ~~~~\forall~ ij \in E.
    \end{align}
The $\mathbf z^\dagger$ is known as an \textit{optimal solution}, and the \textit{value} of the ILP is the value of the objective evaluated at $\mathbf z^\dagger$. 
%The matching $M^\dagger$ constructed from $\mathbf z^\dagger$ is called an optimal match.
Any $\mathbf z$ that satisfies the constraints---but does not necessarily minimize the objective---is simply called a \textit{solution}.  
This problem is an \textit{integer} programming problem as the variables $\mathbf z$ are integer-valued, and is \textit{linear} because both the objective function and the constraints are linear combinations of the $\mathbf{z}$ variables.  

Note that the constraints for LAP ensure that each treated unit is matched to exactly one control unit and \textit{vice versa}.
In other words, every unit under study is covered by exactly one edge.
%MIKE: CITES FOR MINIMUM WEIGHT PERFECT MATCHING PROBLEM
This type of matching is known as a \textit{perfect matching}; hence, LAP is also known as the \textit{minimum cost (or weight) perfect matching problem}. 
%Sanjeewani: do we need to add :"if the objective is to minimizing total dissimilarities.
%5.1.1. heading is not clear.
%MIKE: NEED TO REREAD.
%Sanjeewani : can't see clearly as a new section.

\subsubsection{Solving Integer Linear Programming Problems}
%MIKE: Check this.
There are, broadly speaking, two kinds of approaches for solving these types of ILP matching problems.
The first approach is to work directly on the integer program.
A common technique is to relax the integer constraint on the variables $\mathbf{z}$ to allow $z_{ij}$ to take values within the entire interval $[0,1]$.
%MIKE: NEED CITE
This relaxation results in a standard linear programming (LP) problem, which can be solved in polynomial time~\citep{khachiyan2000integer}.
%MIKE: Also cite Maximum Matching and a 0--1 polytope by Edmonds.
After this relaxation, additional constraints---for example, blossom inequalites~\citep{edmonds1965paths, edmonds1965maximum}---can be iteratively added to the LP to force a solution with 0--1-valued variables.

A particularly interesting instance of the LP relaxation approach occurs when all costs $w_{ij}$ are integer-valued.
In this case, the \textit{integrality theorem}~\citep{dasgupta2008algorithms} ensures the existence of an optimal solution $\mathbf z^\dagger$ to the LP such satisfying $z^\dagger_{ij} \in \{0,1\}$.
Hence, a standard linear program solver---for example, the simplex method~\citep{nelder1965simplex, dantzig1990origins}---can exactly solve the original ILP.
In practice, this is a quite common setting; edge costs are often multiplied by a large power of 10 and rounded to the nearest integer before the optimization problem is initialized.
While this necessarily yields an approximation to the original statistical matching problem, such a matching tends to be acceptable in practice.

Primal-dual methods provide another technique to solve ILP problems.
%MIKE: Do we want to use the word coefficients instead of costs?  I'm leaning towards costs
%Sanjeewani: Coefficient should work better than cost or just roles of the variables are switched enough
In \textit{very} crude terms, the dual of an optimization problem is also an optimization problem, but the roles of the variables the costs are switched and the objective function is ``flipped''---for example, the dual of a minimization problem is a maximization problem~\citep{bachem1992linear}.
%Sanjeewani: Is this " a solution to the maximization problem yields an lower-bound on the objective"
Duality allows for quick computation of both lower and upper bounds to the objective of an optimization problem; for example, a solution to a minimization problem yields an upper bound on the objective, and a solution to the dual of this problem yields a lower-bound on this objective.
%MIKE: NEed cite where reader can find examples.
Additionally, under certain conditions,
%---for example, the Karush–Kuhn–Tucker conditions---
%Sanjeewani: Is this " the value of the optimization problem" or the " the results or final answer for the optimization problem"
the value of the optimization problem and the value of its dual will be the same---a property known as \textit{strong duality}.  
When strong duality holds, an arbitrarily good solution can be found by iteratively switching between the original problem and dual problem, where the solution for the dual problem helps improve the solution for the original optimization problem and \textit{vice versa}~\citep{fang2017extended}.

The second approach is to iteratively manipulate characteristics of the matching graph $G$---for example, edge costs, cycles, minimum cuts, or shortest paths~\citep{kovacs2015minimum}---until an optimal solution is found.
For example, an instance of LAP with $N/2$ treated and control units can be solved using the Hungarian 
algorithm~\citep{kuhn1955hungarian,  munkres1957algorithms, dutta2015note}.  
%Sanjeewani: cost or dissimilarity
This algorithm can be viewed as performing a series of manipulations on the $N/2 \times N/2$ cost matrix $W$---the entry in the $i$th row and $j$th column of $W$ is the cost $w_{ij}$.
For small instances of LAP, these manipulations can be performed using a pen and paper.
We now describe this implementation of the Hungarian algorithm in detail.

\begin{figure}
\begin{center}
\begin{minipage}{2 in}
    \begin{tikzpicture}[scale=1]
	\tikzstyle{every node} = [circle, fill=red!30]
		%Treated nodes
		\node (1) at (1,4) {1};
		\node (2) at (1,3) {2};
		\node (3) at (1,2) {3};
		\node (4) at (1,1) {4};
		\node (5) at (1,0) {5};
		
		\tikzstyle{every node} = [circle, fill=blue!30]
		%Control nodes
		\node (6) at (4,4) {1};
		\node (7) at (4,3) {2};
		\node (8) at (4,2) {3};
		\node (9) at (4,1) {4};
		\node (10) at (4,0) {5};

		\draw (1) -- (6);
		\draw (1) -- (7);
		\draw (1) -- (9);
		\draw (1) -- (10);
		\draw (1) -- (8);
		\draw (2) -- (6);
		\draw (2) -- (7);
		\draw (2) -- (8);
		\draw (2) -- (9);
		\draw (2) -- (10);
		\draw (3) -- (6);
		\draw (3) -- (7);
		\draw (3) -- (8);
		\draw (3) -- (9);
		\draw (3) -- (10);
		\draw (4) -- (6);
		\draw (4) -- (7);
		\draw (4) -- (8);
		\draw (4) -- (9);
		\draw (4) -- (10);
		\draw (5) -- (6);
		\draw (5) -- (7);
		\draw (5) -- (8);
		\draw (5) -- (9);
		\draw (5) -- (10);
	
		%[style = thick][->] (\from) -- (\to);
        \end{tikzpicture}
        
\vspace{.1 in}
\hspace{.7in}
\textbf{(a)}
\end{minipage}
\hspace{2.7in}

\begin{minipage}{2 in}
\begin{tikzpicture}[scale=1]
\centering

\matrix [matrix of nodes, left delimiter=(,right delimiter=) ]
{
$w_{11}$ &$w_{12}$&$w_{13}$&$w_{14}$&$w_{15}$\\
$w_{21}$&$w_{22}$&$w_{23}$&$w_{24}$&$w_{25}$\\
$w_{31}$&$w_{32}$&$w_{33}$&$w_{34}$&$w_{35}$\\
$w_{41}$&$w_{42}$&$w_{43}$&$w_{44}$&$w_{45}$\\
$w_{51}$&$w_{52}$&$w_{53}$&$w_{54}$&$w_{55}$\\
};
\end{tikzpicture}

\centering
\textbf{(b)}
\end{minipage}
\hspace{0.7 in}
  \caption{(a). A complete bipartite graph with $N_T=N_C = 5$. (b). The cost matrix of (a), where a smaller value of $w_{ij}$ indicates that $i$ and $j$ have more similar values of covariates. } 
  \label{fig1}
\end{center}
\end{figure} 

\subsubsection{Hungarian Algorithm for Solving LAP}
The Hungarian algorithm builds an optimal match through selecting entries of the $N/2 \times N/2$ cost matrix $W$~\citep{munkres1957algorithms};
if the entry in the $i$th row and $j$th column of $W$ is selected, then the edge $ij$ is added to the optimal matching $M^\dagger$, and a cost of $w_{ij}$ is incurred.  
%Since LAP aims to find an optimal perfect match, a total of $N/2$ entries must be selected.  
Additionally, from~\citet{kHonig1931grafok}, in order for the match to be perfect, the selected entries must not be coverable by fewer than $N/2$ lines. 

The algorithm works by iteratively adding and subtracting costs from the matrix $W$ to obtain a modified cost matrix $W^\dagger$.  
These operations are performed in such a way to ensure three properties: the optimal solution in $W^\dagger$ is the same as that in $W$; the costs in $W^\dagger$ are non-negative; and that the optimal solution in $W^\dagger$ has a total cost of 0.
Hence, the matching can be verified as optimal through inspection;  it is optimal if and only if the selected entries of $W^\dagger$ are all 0 and cannot be covered by fewer than $N/2$ horizontal or vertical lines.

The algorithm proceeds as follows.  
For brevity, we do not go into detail about how to cover 0 entries with lines in Step~\ref{step3h} or technical proofs as to why repeated applications of Step~\ref{step5h} will lead to convergence of the algorithm.
See~\citet{dutta2015note} for a rigorous discussion.
%For a rigorous proof th 
%For a more rigorous treatment of the Hungarian algorithm with 
% With few exceptions, we employ the terminology and mathematical notation of~\citet{dutta2015note}.

% while ensuring that all entries of the cost matrix are non-negative.
% Recall that LAP is equivalent to finding a minimum cost perfect matching.  
% LAP  adding an edge $ij \in M$ incurs a cost of $w_{ij}$
% that cannot be covered with fewer than $n$ horizontal or vertical lines.
% Given an $n \times n$ cost matrix $W$, LAP amounts to selecting $n$ entries of this matrix such that the sum of the weights are minimized. 
% The intuition behind the Hungarian algorithm is as follows.

 % LAP is one of the most well-studied optimization problems in the Operational Research literature with different approaches. The Hungarian method developed by Kuhn ~\citep{kuhn1955hungarian} was the first systematic approach for finding the optimal solution to a LAP. The computation complexity of the Hungarian algorithm is $O(n^3)$, where $n=|V_T|=|V_C|$, is the order of the weight matrix. With a few exceptions, this paper employs the terminology and mathematical notation of Jayanta Dutta and S.C. Pal \citep{dutta2015note}.

%MIKE: A sequence of pictures could be nice, but not necessary if it takes up too much space.
 \begin{enumerate}
 \item \label{Step0GM} \textbf{(Initialize)} Begin with an
 %an empty assignment or matching, $M_0 = \emptyset$, and an
 $N/2 \times N/2$ cost matrix $W$. 
 
\item \textbf{(Subtract the minimum of each row)} For each row $i$, find %$W_{i\cdot}^{-} = \min_{j} W_{ij}$, 
the smallest entry of $W$ in row $i$. 
Subtract all costs in row $i$ by 
this entry to form a new cost matrix $W^{r}$.
Note,
%$W_{i\cdot}^{-}$ to form a cost matrix $W^{r}$ with entries $W^r_{ij} = W_{ij} - W_{i\cdot}^-$---note that 
$W^r$ will have at least one 0 entry within each row.   
%minimum of each row of the weight matrix from all the weights  of the respective rows. 
\item \textbf{(Subtract the minimum of each column)} 
Similarly, for each column $j$, %find $W_{\cdot j}^{r-} = \min_{i} W^r_{ij}$, 
the smallest entry of $W^r$ in column $j$.
Subtract all costs in column $j$ by 
this entry
%$W_{\cdot j}^{r-}$ 
to form a  cost matrix $W^{c}$.
%with entries $W^c_{ij} = W^r_{ij} - W_{\cdot j}^{r-}$.  
Now, each row and each column of $W^c$ has at least one 0 entry.
 % Modify the resulting matrix by subtracting the minimum weight of each column from all the weights of the respective columns.  

\item \label{step3h} 
%\textbf{(Draw the least number of horizontal and vertical lines to cover all zeroes)} 
\textbf{(Cover all zeroes)} %Perform row and column scanning to 
Cover all zeroes of $W^c$ with as few horizontal and vertical lines as possible.  
Let $L$ denote the total number of lines required.
If $L = N/2$, set $W^\dagger = W^c$ and go to Step~\ref{step7h}. 
Otherwise, proceed to Step~\ref{step4h}.

% draw the mimumum number of horizontal and vertical lines required to pass through all zeroes of $W^c$. 
%Make sure to use the minimum total number of lines to cover all zeroes in the weight matrix.  

\item \label{step4h} \textbf{(Partition entries)} Partition entries of $W^c$ into three components: Those entries that are uncovered by a line $W^{c0}$; those that are covered by exactly one line $W^{c1}$; and those that are covered by two lines $W^{c2}$.  

% Let $l$ be the total number of lines used in step~\ref{step3h} to cover all zeroes.
% If $n=l$, the optimality is reached go to step~\ref{step7h}. Otherwise, proceed to Step~\ref{step5h}.

\item \label{step5h} \textbf{(Find the minimum uncovered cell value)} Find the smallest cost 
of an entry in $W^{c0}$.
%of an uncovered entry $W^{c-}_{ij} \in W^{c0}$.
Subtract this cost from all entries in $W^{c0}$ and add it to all entries in $W^{c2}$ to obtain a new cost matrix $W^{c\prime}$.
Go to Step~\ref{step3h} with $W^c = W^{c\prime}$.

% uncovered entries and add it to all entries that have been covered twice to form a new cost matrix $W_c'$.
% \begin{equation}
%     W_c'
% \end{equation}

% Determine the minimum cell value in the resulting matrix that is not covered by $l$ lines. Subtract this minimum weight from all uncovered weights and add the same weight at the intersection of horizontal and vertical lines. Return to step~\ref{step6h} with the modified weight matrix.

% \item \label{step6h} \textbf{(Repeat until $n=l$)} Repeat Step~\ref{step3h} to ~\ref{step5h} until get $n=l$. If $n=l$, the optimality is reached go to step~\ref{step7h}.

\item \label{step7h} \textbf{(Find optimal matching)} Choose a set of entries $ij$ such that $W^\dagger_{ij} = 0$ for all entries and no entries occur in the same row or column, and let $M^\dagger$ denote this set.  Then, $M^\dagger$ is an optimal matching.

\end{enumerate}

%MIKE: Discussion about runtime of the Hungarian Algorithm.
The Hungarian algorithm requires $O(N^3)$ runtime to terminate.  
The majority of this runtime is devoted to verifying the existence of an optimal solution, for example, for finding optimal matching by drawing the minimum number of lines through the matrix to cover all zeroes.

\subsection{The Linear Unbalanced Assignment Problem}
  The linear unbalanced assignment problem (LUAP) is a extension of LAP in which $N_T < N_C$.  
  The ILP formulation of LUAP is identical to that for LAP except that the constraint~\eqref{constraintC} changes to
  \begin{equation}
    \sum_{j \in V_C} z_{ij} \leq 1~~~~ \forall ~i \in V_T.
  \end{equation}
  %on the var also changes a constraint in the ILP formulation slightly to
  Note that all optimal matching problems are either equivalent to LAP or LUAP.
  
  LUAP can straight-forwardly be reduced to LAP by creating $N_C - N_T$ ``dummy'' treated nodes and setting the cost between these dummy nodes and any control node to be $w^+ = \max_{ij} w_{ij} + 1$.
  This forces an instance where there are an equal number of ``treated'' and control nodes.
  The choice of costs ensures that 
  %the LAP reduction does not prioritize finding a match for a dummy node over one in $V_T$. To be specific, 
  an optimal solution $x^\dagger$ for the original LUAP can be obtained by taking the optimal solution for the LAP reduction and selecting only the $N_T$ variables that are associated with an edge incident to a node $i\in V_T$---swapping one of these edges with one incident to a dummy node will only increase the objective.
  A similar transformation can be performed to prevent certain units from being paired together---that is, between $i' \in V_T$ and $j' \in V_C$ if $i'j' \notin E$.
  In this case, we may set $w^+_{i'j'} = \max_{ij \in E} w_{ij} + 1$ prior to solving the LUAP.

 %MIKE: More info here.  Maybe using results from runtime for the hungarian algorithm.
 %MIKE: NEED CITES.
    Since LUAP can be reduced to LAP, it follows that the Hungarian algorithm can be used to solve instances of LUAP as well.
    However, the addition of dummy nodes may substantially increase the total runtime of the Hungarian algorithm ($O(N^3)$), especially if a large number of dummy nodes are added.  
    Additionally, adding dummy nodes may substantially increase memory requirements---the cost matrix $W$ requires $O(N^2)$ space to store.  
    Thus, attempting to solve LUAP using the Hungarian algorithm may not be an efficient approach matching under big data settings, and historically, other approaches have been used to solve LUAP and optimal matching problems.

\subsection{Maximum Cardinality Matching}

% The statistical matching literature highlights two types of matching techniques: the optimal matching and the greedy matching technique. The optimal matching technique finds the matched sample with the smallest average distance weights,
% while the greedy matching always chooses which element of a set seems to be the best with the lowest weight at the moment. So, the greedy matching selects the closest control match for each treated unit one at a time without reconsidering it at a later point. Hence, the greedy matching approach is not trying to minimize the total distance in the matched sample. Hence, the optimal matching technique is always better than the greedy matching or suboptimal matching technique.

While methods for solving LUAP directly can be implemented to find an optimal matching, in the statistical matching literature, this optimization problem has historically been broken into two separate subproblems:
%MIKE: We haven't defined a "feasible matching"
\begin{enumerate}
    \item \textbf{(Maximum cardinality matching)} Find the maximum cardinality $m^\dagger$ across all possible matchings.
    \item \textbf{(Minimum cost matching)} Find the matching that has the smallest total cost under the constraint that the matching contains $m^\dagger$ matched pairs.
\end{enumerate}
% Note, if $|V_T| \leq |V_C|$ and each treated unit is eligible to be matched to every control unit---\textit{i.e.}~if $G$ is a complete bipartite graph---then the maximum cardinality $m^\dagger = |V_T|$.
We now describe these subproblems in detail, beginning with maximum cardinality matching.

% Recall that a 1:1 matching without replacement $M$ on a graph 
% For any $G = ((V_T, V_C), E)$ is a subset $M\subset E$ such that each treated node $i \in V_T$ and control node $j \in V_C$ is incident with at most one edge in $M$. 
% the maximum cardinality matching problem (MaxCard) is to find a matching in $G$ such that the cardinality of the matching $|M|$ is as large as possible.
% %$Max \{|M| | M \in \mathbb M\}$,  where $\mathbb M$ is all possible match in $G$.

For any bipartite graph $G = ((V_T, V_C), E)$, the maximum cardinality matching problem (MaxCard) is to find a matching in $G$ such that the cardinality of the matching $|M|$ is as large as possible.
MaxCard may be formulated as an ILP where the aim is to
% \textit{integer linear programming} problem (ILP). 
% This formulation associates each edge $ij \in E$ with a binary variable $z_{ij}$, where $z_{ij}=1$ if the match $ij \in M $ and $z_{ij}=0$ if $ij \notin M$.
% MaxCard aims to find, 
% Across all %possible vectors of \textit{variables} 
% variables $\mathbf z = (z_{ij})_{ij \in E}$, an optimal vector 
find an optimal solution $\textbf{z}^\dagger$ satisfying
%that satisfies
$$
        \mathbf z^\dagger = \argmax_{\mathbf z} \sum_{ij\in E} z_{ij}
$$
under the constraints that
\begin{align}
        \sum_{i \in V_T} z_{ij} \leq 1&~~~~ \forall ~j \in V_C, \nn
        \sum_{j \in V_C} z_{ij} \leq 1&~~~~ \forall ~i \in V_T, \nn
        z_{ij} \in \{0,1\}& ~~~~\forall~ ij \in E.
    \end{align}
% This problem is an \textit{integer} programming problem as the variables $\mathbf z$ are integer-valued, and is \textit{linear} because both the objective function and the constraints are linear combinations of the $\mathbf{z}$ variables.  
Note, the matching $M^\dagger$ induced by such a $z^\dagger$ satisfies $|M^\dagger| = m^\dagger$.

Some available matching methods work directly on this objective.  
One notable example, \textit{cardinality matching}~\citep{zubizarreta2014matching}, solves this ILP with an additional constraint ensuring that, for example, the differences in sample means of confounding covariates between treated and matching control groups are within some pre-specified tolerance threshold.
After finding a matching $M^\dagger$ that maximizes the cardinality, 
%this method then finds 
an optimal matching
%~\citep{rosenbaum1989optimal} 
on all units incident to an edge in $M^\dagger$ is obtained before estimating treatment effects.

A traditional approach for solving MaxCard is to first transform this problem into a network flow problem.
Algorithms designed to find maximum flows can then be applied to solve the original MaxCard problem.
These maximum flow algorithms are often computationally efficient, and thus, may be scalable to large observational studies.  
We now describe a solution using this approach---the Ford-Fulkerson algorithm---in detail~\citep{ford1957simple}.

\subsubsection{\label{sec:fordfulk} Ford-Fulkerson for Solving MaxCard}

\begin{figure}
\begin{center}
\begin{minipage}{2in}
    \begin{tikzpicture}[scale=1.15]
		\tikzstyle{every node} = [circle, fill=red!30]
		%Treated nodes
		\node (1) at (1,4) {1};
		\node (2) at (1,3) {2};
		\node (3) at (1,2) {3};
		\node (4) at (1,1) {4};
		\node (5) at (1,0) {5};
		
		\tikzstyle{every node} = [circle, fill=blue!30]
		%Control nodes
		\node (6) at (4,4) {1};
		\node (7) at (4,3) {2};
		\node (8) at (4,2) {3};
		\node (9) at (4,1) {4};
		\node (10) at (4,0) {5};

		\draw (1) -- (6);
		\draw (1) -- (7);
		\draw (1) -- (9);
		\draw (1) -- (10);
		\draw (1) -- (8);
		\draw (2) -- (6);
		\draw (2) -- (7);
		\draw (2) -- (8);
		\draw (2) -- (9);
		\draw (2) -- (10);
		\draw (3) -- (6);
		\draw (3) -- (7);
		\draw (3) -- (8);
		\draw (3) -- (9);
		\draw (3) -- (10);
		\draw (4) -- (6);
		\draw (4) -- (7);
		\draw (4) -- (8);
		\draw (4) -- (9);
		\draw (4) -- (10);
		\draw (5) -- (6);
		\draw (5) -- (7);
		\draw (5) -- (8);
		\draw (5) -- (9);
		\draw (5) -- (10);
	
		%[style = thick][->] (\from) -- (\to);
\end{tikzpicture}

\centering
\textbf{(a)}
\end{minipage}
\hspace{.7in}
\begin{minipage}{2in}
\begin{tikzpicture}[scale=0.9]
    
    \tikzstyle{every node} = [circle, fill=green!30]
		%Sourse node
		\node (0) at (7, 2) {s};
		
		\tikzstyle{every node} = [circle, fill=red!30]
		%Treated nodes
		\node (1) at (9,4) {1};
		\node (2) at (9,3) {2};
		\node (3) at (9,2) {3};
		\node (4) at (9,1) {4};
		\node (5) at (9,0) {5};
		
		\tikzstyle{every node} = [circle, fill=blue!30]
		%Control nodes
		\node (6) at (12,4) {1};
		\node (7) at (12,3) {2};
		\node (8) at (12,2) {3};
		\node (9) at (12,1) {4};
		\node (10) at (12,0) {5};
		
		\tikzstyle{every node} = [circle, fill=green!30]
		\tikzstyle{arrow} = [thick,->,>=stealth]
		%Sourse node
		\node (11) at (14, 2) {t};
	
        \draw [arrow] (0) -- (1);
		\draw [arrow](0) -- (2);
		\draw [arrow](0) -- (3);
		\draw [arrow](0) -- (4);
		\draw [arrow](0) -- (5);
		\draw [arrow] (1) -- (6);
		\draw [arrow](1) -- (7);
		\draw [arrow](1) -- (8);
		\draw [arrow](1) -- (9);
		\draw [arrow](1) -- (10);
		\draw [arrow](2) -- (6);
		\draw [arrow](2) -- (7);
		\draw [arrow](2) -- (8);
		\draw [arrow](2) -- (9);
		\draw [arrow](2) -- (10);
		\draw [arrow](3) -- (6);
		\draw [arrow](3) -- (7);
		\draw [arrow](3) -- (8);
		\draw [arrow](3) -- (9);
		\draw [arrow](3) -- (10);
		\draw [arrow](4) -- (6);
		\draw [arrow](4) -- (7);
		\draw [arrow](4) -- (8);
		\draw [arrow](4) -- (9);
		\draw [arrow](4) -- (10);
		\draw [arrow](5) -- (6);
		\draw [arrow](5) -- (7);
		\draw [arrow](5) -- (8);
		\draw [arrow](5) -- (9);
		\draw [arrow](5) -- (10);
		\draw [arrow] (6) -- (11);
		\draw [arrow](7) -- (11);
		\draw [arrow](8) -- (11);
		\draw [arrow](9) -- (11);
		\draw [arrow](10) -- (11);
		
		%[style = thick][->] [arrow] (\from) -- (\to);
\end{tikzpicture}

\centering
\textbf{      (b)}
\end{minipage}
\hspace{0.7 in}
	
\caption{(a). A bipartite graph with five treated and control units. (b). The network flow graph for (a). } \label{fig2}
\end{center}
\end{figure}
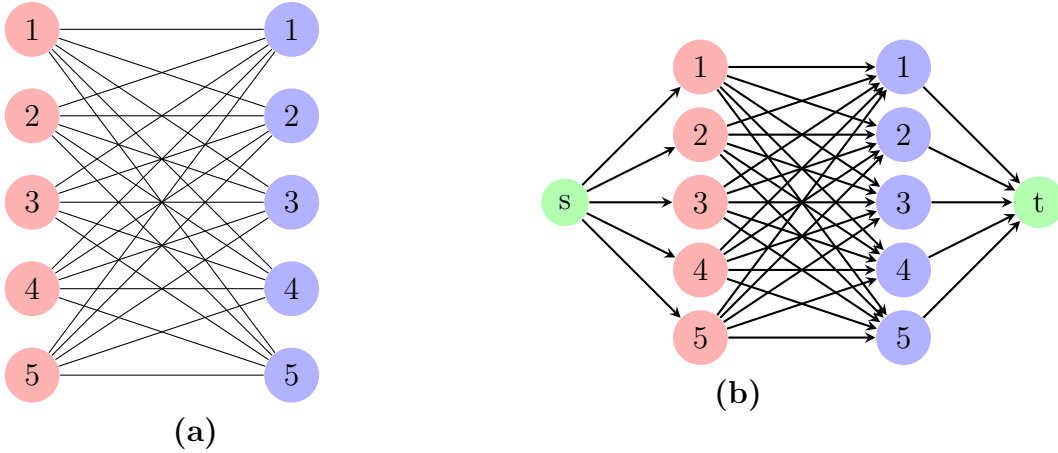

%A traditional approach to solving 
We begin by reducing MaxCard to a maximum flow problem.
To do this, we first transform $G$ to a digraph $G' = (V', E')$---that is, each edge in $E'$ is now directed.
Specifically, we allow edges in $E'$ to travel from a node in $V_T$ to a node in $V_C$, but not the other direction: for $i \in V_T$, $j \in V_C$, and $ij \in E$, we have $\vec{ij} \in E'$, but $\vec{ji} \notin E'$.
We then add a \textit{source} node $s$ and a \textit{sink} node $t$ to $G$, and we connect these nodes to $G'$ by adding edges traveling from the source node to each node in $V_T$ and edges traveling from each node in $V_C$ to the sink node: for $i \in V_T$, $\vec{si} \in E'$, and for $j \in V_C: \vec{jt} \in E'$.
%%%Dr.Higgins, please look at this
Finally, we assign each edge in $e \in E'$ a \textit{capacity} $c_{e}$ equal to 1.
Figure~\ref{fig2} details this transformation.

% we create  and a and a ``sink'' node into the graph $G$ and assigning all edges within $G$ to have a ``capacity'' of 1. 
% are introduced into the graph , and algorithms designed to maximize the ``flow'' from the source to the sink are applied.  
% Then, algorithms for finding the maximum flow---for example, the Ford-Fulkerson algorithm---

% Consider $G$ is a bipartite digraph, with $\vec{ij}\in E$ being the edge set of $G$. Let a source $s$ and a sink $t$ be two new vertices, and there will be a positive capacity $c_{ij}$ on every edge $\vec{ij}\in E$. Construct a graph $G'$ with vertex set $V \cup \{s,t\}$ where $c_{si}=1$ for every $i\in V_T$ and  $c_{jt}=1$ for every $j\in V_C$. The graph $G'$ is called a network flow graph and shows in Figure 2.1 (b).

A \textit{flow} on the digraph $G'$ from the source $s$ to the sink $t$ is %defined as 
a real-valued function $f$ on each edge $e \in E'$ satisfying the following conditions:
\begin{enumerate}
 \item For any edge $e \in E'$ : $0 \leq f(e) \leq c_{e}$.  
 If $f(e) = c_{e}$, we say that the flow is \textit{saturated} on that edge.
 \item For any node $j\in V' \backslash \{s,t\}$,  the total flow into the node $j$ is same as the total flow out of the node.  
 That is,
\begin{equation} 
    \sum_{i:\vec{ij} \in E'}f(\vec{ij}) =\sum_{k:\vec{jk}\in E'}f(\vec{jk}).
    \label{eq:defineflow}
\end{equation}    
\end{enumerate}
The value $|f|=\sum_{i: \vec{si} \in E'}f(\vec{si})$ is the \textit{total flow} out from the source $s$, and hence, from~\eqref{eq:defineflow}, the total flow entering into $t$ is $|f|$.
%MIKE Defined matching induced by a flow here.  Check the next two paras
Under this setup, each flow $f$ on $G'$ \textit{induces} a matching $M_f \subset E$ obtained by selecting the edges that the flow saturates:
\begin{equation}
    M_f = \left\{ij \in E: i \in V_T, j \in V_C, f(\vec{ij}) = 1\right\}.
\end{equation}

The maximum flow problem is to find a flow $f^\dagger$ that maximizes the total flow into $t$. 
If all capacities are integers---as is the case with MaxCard---it is possible to find such an $f^\dagger$ with integer values for all edges: $f(e) \in \mathbb N \cup 0~~\forall~e \in E'$~\citep{dasgupta2008algorithms}.
Upon finding such a maximum flow $f^\dagger$, a maximum cardinality matching $M^\dagger$ is a matching induced by this flow: $M^\dagger = M_{f^\dagger} $.

%MIKE: Please check to see if this is correct
%MIKE: Need a cite to ford fulkerson.
The Ford-Fulkerson algorithm (FFA) is commonly used to solve maximum flow problems.
Intuitively, FFA works by starting from an initial flow $f$ and iteratively finding paths of edges in $G'$ from $s$ to $t$ that will lead to increases in the total flow of $f$.
%flow $f$ will increase the total flow.  
%MIKE: Can do another cite here.  
%For solving MaxCard, FFA is guaranteed to find the maximum flow in, at most, $V_T$ iterations.

FFA is most easily described through the introduction of residual graphs.
Given the maximum flow digraph $G' = (V',E')$ and a flow $f$, the \textit{residual} graph $H=(V', E_f)$ is a digraph on the nodes $V$.
For each edge $\vec{ij} \in E'$, there is a  ``forwards'  $\vec{ij}$ and ``backwards''  $\vec{ji}$ version of this edge in the residual edge set $E_f$:
\begin{equation}
    E_f = \left\{\vec{ij} \cup \vec{ji} : \vec{ij} \in E'\right \}
\end{equation}
%MIKE: Making this for Max Card.  In general, change 1 to c_{ij}
For MaxCard in particular, forward edges $\vec{ij}$ have a \textit{residual capacity} of $\delta(\vec{ij}) = 1 - f(\vec{ij})$, which denotes the unused capacity on edge $\vec{ij}$. 
Backward edges $\vec{ji}$ have capacity $\delta(\vec{ji}) = f(\vec{ij})$, which denotes how much the flow on edge $\vec{ij}$ can be suppressed.
%be reduced when preventing the flow from entering edge $\vec{ij}$. 
That is,
\begin{equation}
    \label{eq:definecapresid}
    \delta(\vec{ij}) = \left\{ \begin{array}{ll}
         1 - f(\vec{ij}), & \vec{ij} \in E', \\
         f(\vec{ij}), & \vec{ji} \in E'.
    \end{array}\right.
\end{equation}

%MIKE: Need to fix some of the vecs here.
Once the residual graph $H$ is constructed, FFA finds paths  $P = \left\{\vec{si_1}, \overrightarrow{i_1i_2}. \ldots, \overrightarrow{i_{\ell -1}i_{\ell}}, \vec{i_{\ell}t}\right\}$ from $s$ to $t$ within this residual graph such that the capacity $\delta(\vec{ij}) > 0$ for each edge $\vec{ij} \in P$.
These paths are called \textit{augmenting paths}.
The current flow $f$ can then be improved 
%An improved flow $f$ can then be obtained 
by adding flow to the forward edges and decreasing flow to the backwards edges along this path.  

Rigorously, FFA for MaxCard is performed as follows:
%Inputs : Given a graph $G=(V,E)$ with flow capacities $c_{ij} ; ij \in E$, a source vertex $s$ and a sink vertex $t$.
\begin{enumerate}
\item \textbf{(Initialize flow)} Set $f(\vec{ij}) = 0$ for all edges $\vec ij \in E'$.
\item \label{step2ffa} \textbf{(Update residual graph)} Update the residual graph $H = (V, E_f)$ with capacities given in~\eqref{eq:definecapresid}.
\item \textbf{(Find augmenting path or terminate)} 
    \label{step3ffa}
    Find an augmenting path $P = \left\{\vec{si_1}, \overrightarrow{i_1i_2}. \ldots, \overrightarrow{i_{\ell -1}i_{\ell}}, \vec{i_{\ell}t}\right\}$ from $s$ to $t$ such that $\delta(\vec{ij}) = 1$ for all edges $\vec{ij} \in P$.
    
    If no such path exists, stop.  
    %while thee exist some s-t path $H$ in residual graph, such that $\delta(H)>0$ for all edges $ij \in H$, augment along $H$ by $\delta(H)$ using one of the following way,
    \item \textbf{(Augment the flow)} Update the flow $f$
    %MIKE: Particular for MaxCard
    along all edges $\vec{ij} \in P$:
    %as follows:
    %\begin{enumerate}
    %MIKE: Particular for MaxCard
%        \item Find $\delta_{min} = \min\{\delta(\vec{ij}): \vec{ij} \in P\}$.
 %       \item Update the flow $f$ along all edges $\vec{ij} \in P$:
        \begin{equation}
            \left. \begin{array}{ll}
                f(\vec{ij})  \longleftarrow f(\vec{ij})+1, & \vec{ij} \in P,~ \vec{ij} \in E',\\
               % f(\vec{ji})  \longleftarrow f(\vec{ji})-1,&  \vec{ij} \in P,~ \vec{ji} \in  E'.\\
                 f(\vec{ij})  \longleftarrow
                f(\vec{ij})-1,&  \vec{ji} \in P,~ \vec{ij} \in  E'.\\
            \end{array}\right.
        \end{equation}
        % \begin{itemize} 
        %     \item If $\vec{ij} \in E'$, update $f(\vec{ij})  \longleftarrow f(\vec{ij})+\delta_{min}$.
        %     %Update the capacities $\delta(\vec{ij}) \longleftarrow \delta(\vec{ij})-\delta_{min}$ and $\delta(\vec{ji}) =  \delta$.  
        %     \item If $\vec{ij} \notin E'$, update $f(\vec{ji})  \longleftarrow f(\vec{ji})-\delta_{min}$. 
        %     %and $\delta(\vec{ij}) \longleftarrow \delta(\vec{ij})+\delta_{min}$.
        % \end{itemize}
%    \end{enumerate}
    Return to Step~\ref{step2ffa}.
\end{enumerate}
% \begin{itemize}
% \item  
% \item  $f_{ij} \longleftarrow f_{ij}-\delta(H)$; use the edge in the reverse direction which is actually leads to decrements in the flow that already have.
% \end{itemize}
%\end{enumerate}

%MIKE: Need cites.
Each iteration of FFA increases the flow of $f$ by $1$. 
For general maximum flow problems, finding an augmenting path takes $O(|E'|)$ time.
Moreover, if all capacities in the maximum flow problem are integer-valued, then the flow $f$ at termination in Step~\ref{step3ffa} is a maximum flow, and reaching this flow requires, at most, $|f^\dagger|$ iterations.
%Morofind a maximum flow $f^\dagger$ in, at most, $|f^\dagger|$ iterations.
In particular, for MaxCard, the maximum cardinality $m^\dagger \leq |V_T| < N$, and so, total runtime of FFA is bounded by $O(N|E|) \leq O(N^3)$.
Moreover, when the graph is \textit{sparse}---that is, when the number of edges is proportional to the number of nodes---this runtime is reduced to $O(N^2)$.
In practice, FFA tends to be computationally efficient enough for most statistical matching applications.

\begin{figure}
\begin{center}
\begin{minipage}{2 in}
\begin{tikzpicture}[scale=0.8] 
		%Sourse node
		\node (0) [circle, fill=green!30] at (-1, 3) {s};
		%Treated nodes
		\node (1)[circle, fill=red!30] at (1,4) {$T_1$};
		\node (2)[circle, fill=red!30] at (1,2) {$T_2$};

		%Sourse node
		\node (5) [circle, fill=green!30] at (6, 3) {t};
		\tikzstyle{arrow} = [thick,->,>=stealth]
		%Control nodes
		\node (3)[circle, fill=blue!30] at (4,4) {$C_1$};
		\node (4) [circle, fill=blue!30]at (4,2) {$C_2$};

\draw (1) edge[arrow, above] node{$(1, 0)$} (3)
(1) edge[arrow,above][style = ultra thick, color = red]node[pos=0.4]{$(1,1)$} (4)
%(2) edge[arrow,below] node[pos=0.3]{$(,0)$} (3)
(2) edge[arrow, below] node{$(1,0)$} (4)
(0) edge[arrow,above][style = ultra thick, color = red]node[pos=0.3]{$(1,1)$} (1)
(0) edge[arrow,below]node[pos=0.3]{$(1,0)$} (2)
(3) edge[arrow,above]node[pos=0.5]{$(1,0)$} (5)
(4) edge[arrow,below][style = ultra thick, color = red]node[pos=0.5]{$(1,1)$} (5);
%\vspace{0.2in}
		%[style = thick][->] (\from) -- (\to);
\end{tikzpicture}\\ 
\centering
\textbf{(a)} 
\end{minipage}
\hspace{0.7 in}
\begin{minipage}{2 in}
\begin{tikzpicture}[scale=0.8]

			%Sourse node
		\node (0) [circle, fill=green!30] at (7, 3) {s};
		%Treated nodes
		\node (1)[circle, fill=red!30]  at (9,4) {$T_1$};
		\node (2)[circle, fill=red!30]  at (9,2) {$T_2$};

        	%Sourse node
		\node (5) [circle, fill=green!30] at (14, 3) {t};
		%Control nodes
		\node (3) [circle, fill=blue!30] at (12,4) {$C_1$};
		\node (4) [circle, fill=blue!30] at (12,2) {$C_2$};

\tikzstyle{arrow} = [thick,->,>=stealth]
	\draw 
(1) edge[arrow, above,color=blue] node{$(1)$} (3)
(1) edge[arrow,above, bend left=10]node[pos=0.5]{$(0)$} (4)
%(2) edge[arrow,above] node[pos=0.1]{$\delta=1$} (3)
(2) edge[arrow , below, color=blue] node{$(1)$} (4)
(5) edge[arrow, below, bend left=10] node{$(1)$} (4)
(4) edge[arrow,below,bend left=10,color=blue] node{$(1)$}(1)
(1) edge[arrow, below, bend left=10] node{$(1)$} (0)
(0) edge[arrow,above,  bend left=10]node[pos=0.3]{$(0)$} (1)
(0) edge[arrow,below,color=blue]node[pos=0.3]{$(1)$} (2)
(3) edge[arrow,above,color=blue]node[pos=0.4]{$(1)$} (5)
(4) edge[arrow,above, bend left=10]node[pos=0.3]{$(0)$} (5);;

		%[style = thick][->] [arrow] (\from) -- (\to);

\end{tikzpicture}\\
\centering
\textbf{(b)} 
\end{minipage}
\end{center}
%%%%%%%%%%%%%%%%%%%%%%%%%%%%%%%%%%%%%%%%%%%%%%%%%%%%%%%%%
\begin{center}
\begin{minipage}{2 in}
\begin{tikzpicture}[scale=0.8] 
		%Sourse node
		\node (0) [circle, fill=green!30] at (-1, 3) {s};
		%Treated nodes
		\node (1)[circle, fill=red!30] at (1,4) {$T_1$};
		\node (2)[circle, fill=red!30] at (1,2) {$T_2$};

		%Sourse node
		\node (5) [circle, fill=green!30] at (6, 3) {t};
		\tikzstyle{arrow} = [thick,->,>=stealth]
		%Control nodes
		\node (3)[circle, fill=blue!30] at (4,4) {$C_1$};
		\node (4) [circle, fill=blue!30]at (4,2) {$C_2$};

\draw (1) edge[arrow, above][style = ultra thick, color =red] node{$(1, 1)$} (3)
(1) edge[arrow,above]node[pos=0.5]{$(1,0)$} (4)
%(2) edge[arrow,below] node[pos=0.3]{$(,0)$} (3)
(2) edge[arrow, below][style = ultra thick, color =red] node{$(1,1)$} (4)
(0) edge[arrow,above][style = ultra thick, color = red]node[pos=0.3]{$(1,1)$} (1)
(0) edge[arrow,below][style = ultra thick, color =red]node[pos=0.3]{$(1,1)$} (2)
(3) edge[arrow,above][style = ultra thick, color =red]node[pos=0.5]{$(1,1)$} (5)
(4) edge[arrow,below][style = ultra thick, color = red]node[pos=0.5]{$(1,1)$} (5);
%\vspace{0.2in}
		%[style = thick][->] (\from) -- (\to);
\end{tikzpicture}\\ % pic 1
\centering
\textbf{(c)}
\end{minipage}
\hspace{0.7 in}
\begin{minipage}{2 in}
\begin{tikzpicture}[scale=0.9]

			%Sourse node
		\node (0) [circle, fill=green!30] at (7, 3) {s};
		%Treated nodes
		\node (1)[circle, fill=red!30]  at (9,4) {$T_1$};
		\node (2)[circle, fill=red!30]  at (9,2) {$T_2$};

        	%Sourse node
		\node (5) [circle, fill=green!30] at (14, 3) {t};
		%Control nodes
		\node (3) [circle, fill=blue!30] at (12,4) {$C_1$};
		\node (4) [circle, fill=blue!30] at (12,2) {$C_2$};

\tikzstyle{arrow} = [thick,->,>=stealth]
	\draw 
(1) edge[arrow, above,bend left=10][pos=0.5] node{$(0)$} (3)
(3) edge[arrow, below,bend left=10][pos=0.5] node{$(1)$} (1)
(1) edge[arrow,above]node[pos=0.8]{$(1)$} (4)
%(2) edge[arrow,above] node[pos=0.1]{$\delta=1$} (3)
(2) edge[arrow ,above, bend left=10][pos=0.2] node{$(0)$} (4)
(4) edge[arrow,below, bend left=10]node[pos=0.5]{$(1)$} (2)
(5) edge[arrow, below, bend left=10][pos=0.5] node{$(1)$} (4)
(4) edge[arrow,above, bend left=10] node[pos=0.2]{$(0)$}(5)

%(4) edge[arrow,below,bend left=10] node{$(1)$}(1)
(1) edge[arrow, below, bend left=10][pos=0.2] node{$(1)$} (0)
(0) edge[arrow,above,  bend left=10]node[pos=0.5]{$(0)$} (1)
(0) edge[arrow,above,bend left=10][pos=0.8]node{$(0)$} (2)
(2) edge[arrow,below,bend left=10]node[pos=0.5]{$(1)$} (0)
(3) edge[arrow,above,bend left=10]node[pos=0.5]{$(0)$} (5)
(5) edge[arrow,below,bend left=10]node[pos=0.8]{$(1)$} (3)
;;

		%[style = thick][->] [arrow] (\from) -- (\to);

\end{tikzpicture}\\
\centering
\textbf{(d)}
\end{minipage}
\end{center}
	\caption{(a). The flow network $G$ and initial flow $f$ with (capacity, flow). (b) The residual graph for (a) with augmenting path $p$ in blue and residual capacity ($\delta$); Consider the reverse-path $C_2-T_1$ and selecting path $s-T_2-C_2-T_1-C_1-t$. (c). The flow in $G$ that results from augmenting along path $p$ by its residual capacity. (d). The residual network induced by the flow in (c); no path can be found  $s-t$ with all edges those with $\delta=1$.} \label{fig3}
	
\end{figure}
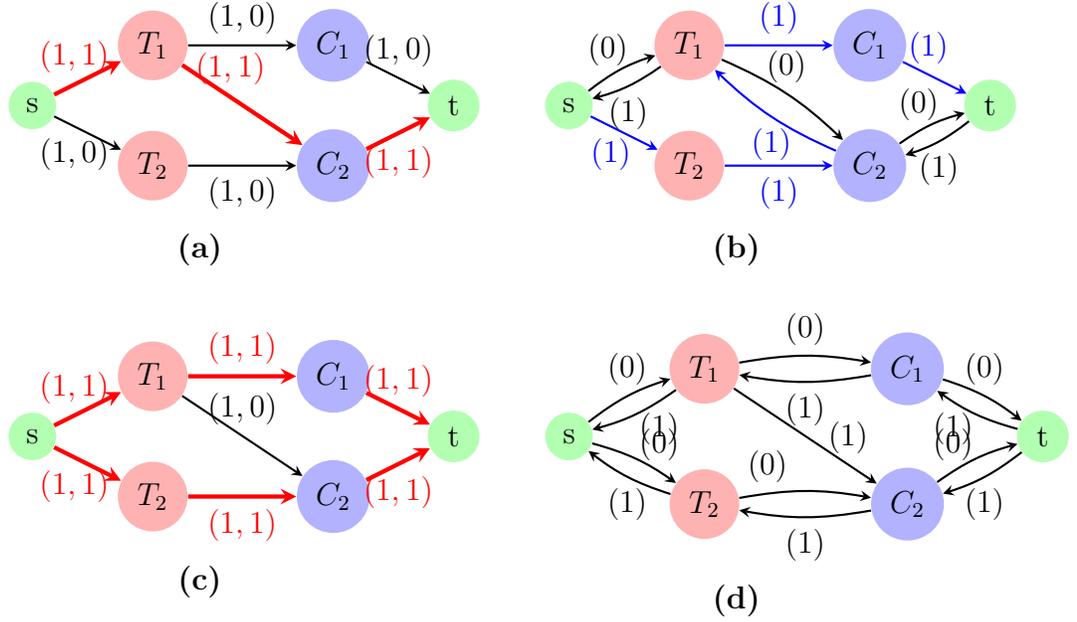

\subsection{Minimum Cost Matching}
Recall that, in the matching graph $G = (V, E)$, each edge $ij \in E$ has a cost $w_{ij} \geq 0$.
The general form of a minimum cost matching problem (MinCost) is to find a matching $M^\dagger$ that minimizes the total cost~\eqref{eq:totalcost} under a constraint that $M^\dagger$ has sufficiently large cardinality.
Constraints on the cardinality of the matching prevent a trivial optimal solution of $M^\dagger$ containing no matched pairs.

As with MaxCard, MinCost can be formulated as an ILP.
For any size of matching $m$, we aim to find an optimal solution $z^\dagger$ satisfying
$$
        \mathbf z^\dagger = \argmax_{\mathbf z} \sum_{ij\in E} w_{ij}z_{ij}
$$
under the constraints that
\begin{align}
        \sum_{i \in V_T} z_{ij} \leq 1~~~~ \forall ~j \in V_C, \nn
        \sum_{j \in V_C} z_{ij} \leq 1~~~~ \forall ~i \in V_T, \nn
        \sum_{i \in V_T}\sum_{j \in V_C} z_{ij} \geq m, \nn
        z_{ij} \in \{0,1\} ~~~~\forall~ ij \in E.
    \end{align}
The constraint $\sum_{i \in V_T}\sum_{j \in V_C} z_{ij} \geq m$ ensures that the optimal matching $M^\dagger$ satisfies $|M^\dagger| \geq m$ (and, in fact, $|M^\dagger| = m$, as any extra edges in $M^\dagger$ can be removed without an increase in the total cost). 
In practice, optimal matching problems will set the cardinality to $m^\dagger$, the maximum cardinality possible for a match.

\subsubsection{Cycle Canceling for Solving MinCost}
%MIKE: A summary of other approaches here.  Ideally, one or two sentences each.
%MIKE: TO INCLUDE

%%%%%%%%%%%%%%%%%%%%%%%%%%%%%%%%%%%%%%%%%%%%%%%%%
Apart from the linear programming approach, there are a variety of approaches for solving MinCost.
We discuss one of these approaches---cycle canceling---while 
%with the Bellman-Ford algorithm---while 
noting that other approaches, including cost-scaling, relaxation, and simplex approaches, may also yield relatively efficient solutions for MinCost. 

As with MaxCard, cycle-canceling approaches for solving MinCost begin by transforming the problem into an optimal flow problem.
The digraph $G' = (V,E')$ described in Section~\ref{sec:fordfulk} is constructed.
For completeness, costs $w$ are defined on all edges $\vec{ij} \in E'$ by setting $w_{si} = 1$ for all $i \in V_T$ and $w_{jt} = 1$ for all $j \in V_C$.
FFA approaches can then be used to find an initial flow $f_0$ satisfying $|f_0| = m$.
Finally, the residual graph $H = (V, E_{f_0})$ is constructed, and
costs $w^{H}$ are assigned to each edge $\vec{ij} \in E_{f_0}$ as follows:
\begin{equation}
    \label{eq:definecostsresid}
    w^{H}_{ij} = \left\{
    \begin{array}{ll}
         w_{ij},& \vec{ij} \in E'~~\text{and}~~f_0(\vec{ij}) = 1, \\
         w_{ij},& \vec{ji} \in E'~~\text{and}~~f_0(\vec{ji}) = 0,\\
         -w_{ij}, & \textit{otherwise}.
    \end{array}
    \right.
\end{equation}
That is, costs are positive for forward edges that are used the flow from $s$ to $t$ and for backwards edges not used in this flow; costs are negative otherwise.

%MIKE: USE CAPITAL Y for cycle, to not confuse with cycle.
Searching negative cycles and canceling them with a cycle canceling algorithm will then find the minimum cost for the matching.  
%A minimum cost matching can then be obtained by cycle canceling algorithms search for negative cycles.
A \textit{cycle} $C$ in the residual graph $H$ is a path that begins and ends at the same node $C =  \left\{\overrightarrow{i_1i_2}, \overrightarrow{i_2i_3}. \ldots, \overrightarrow{i_{\ell -1}i_{\ell}}, \overrightarrow{i_{\ell}i_1}\right\}$.
A \textit{negative cycle} is a cycle $C^-$ in which the sum of the costs along edges in the cycle is negative: $\sum_{\vec{ij} \in C^-} w_{ij} < 0$.
It can be shown that the matching $M$ induced by a flow $f$ is a minimum cost matching if and only if there are no negative cycles within the corresponding residual graph \citep{klein1967primal}.
%MIKE: Need Cites.
There are a variety of methods for finding negative cycles, including the Bellman-Ford algorithm and minimum-mean cycle approaches.

%MIKE: Description of how to go from a negative cycle to update the flow (and thus the matching).
For MinCost specifically, each cycle within the residual graph will have the same number of forward edges traveling from a treated unit to a control unit as backward edges traveling from a control unit to a treated unit. 
Once a negative cycle $C^-$ is found, the flow is updated by pushing flow forward through the backward edges in $C^-$ and preventing flow from traveling through the forward edges in $C^-$.
The matching induced by the updated flow will have the same cardinality as the matching with the original flow but will have a smaller total cost. 
The process of updating the flow from the negative cycle is called \textit{cycle canceling}.

Rigorously, cycle canceling for MinCost is performed as follows:
%Inputs : Given a graph $G=(V,E)$ with flow capacities $c_{ij} ; ij \in E$, a source vertex $s$ and a sink vertex $t$.
\begin{enumerate}
\item \textbf{(Initialize flow)} Find initial flow $f$ on $G' = (V,E')$ with total flow $|f| = m$.  
Define costs $w$ on all edges $E'$ as previously described.
\item \label{step2cc} \textbf{(Update residual graph)} Update the residual graph $H = (V, E_f)$ with costs given in~\eqref{eq:definecostsresid}.
\item \textbf{(Find negative cycle or terminate)} 
    \label{step3ffa}
    Find a cycle $C^- = \left\{\overrightarrow{i_1i_2}, \overrightarrow{i_2i_3}. \ldots, \overrightarrow{i_{\ell -1}i_{\ell}}, \overrightarrow{i_{\ell}i_1}\right\}$ satisfying
    $\sum_{\vec{ij} \in C^-} w_{ij} < 0$.
    
    If no such cycle exists, stop.  
    %while thee exist some s-t path $H$ in residual graph, such that $\delta(H)>0$ for all edges $ij \in H$, augment along $H$ by $\delta(H)$ using one of the following way,
    \item \textbf{(Update the flow)} Update the flow $f$
    %MIKE: Particular for MaxCard
    along all edges $\vec{ij} \in C$ as follows:
    %as follows:
    %\begin{enumerate}
    %MIKE: Particular for MaxCard
%        \item Find $\delta_{min} = \min\{\delta(\vec{ij}): \vec{ij} \in P\}$.
 %       \item Update the flow $f$ along all edges $\vec{ij} \in P$:
        \begin{equation}
            \left. \begin{array}{ll}
                f(\vec{ij})  \longleftarrow f(\vec{ij})+1, &  \vec{ij} \in C^-,~ \vec{ij} \in  E',\\
                f(\vec{ij})  \longleftarrow f(\vec{ij})-1, & \vec{ji} \in C^- , ~ \vec{ij} \in E'.
            \end{array}\right.
        \end{equation}
        % \begin{itemize} 
        %     \item If $\vec{ij} \in E'$, update $f(\vec{ij})  \longleftarrow f(\vec{ij})+\delta_{min}$.
        %     %Update the capacities $\delta(\vec{ij}) \longleftarrow \delta(\vec{ij})-\delta_{min}$ and $\delta(\vec{ji}) =  \delta$.  
        %     \item If $\vec{ij} \notin E'$, update $f(\vec{ji})  \longleftarrow f(\vec{ji})-\delta_{min}$. 
        %     %and $\delta(\vec{ij}) \longleftarrow \delta(\vec{ij})+\delta_{min}$.
        % \end{itemize}
%    \end{enumerate}
    Return to Step~\ref{step2cc}.
\end{enumerate}

%MIKE: ALso find the Karp paper and cite it here.
As mentioned before, each iteration of the cycle canceling algorithm will find a flow with the same total flow but a smaller total cost.
Standard approaches for finding negative cycles require $O(N|E'|)$ time ~\citep{goldberg1989finding}.  
However, unlike with MaxCard, there may not be a restrictive upper bound for the number of iterations required to find an optimal solution.
If all costs are integer-valued, cycle canceling algorithms can terminate in $O(N|E'|\sum_{\vec{ij} \in E'} w_{ij})$ iterations as each iteration will reduce the total cost by at least 1~\citep{kovacs2015minimum}.
%MIKE: Cite Tarjan and whatnot
%MIKE: Need to doublecheck runtime.
Additionally, some algorithms have been developed for MinCost that are guaranteed to terminate in polynomial time with respect to $N$, even if costs are not integer-valued.
The most well-known of these algorithms, minimum mean-cycle cancelling~\citep{goldberg1989finding, radzik1994tight}, requires at most $O(N|E'|^2)\leq O(N^5)$ iterations, 
leading to a total runtime of $O(N^2|E'|^3) \leq O(N^8)$.
This is substantially more computationally complex than FFA.
Again, ensuring sparsity in the matching graph can dramatically reduce the runtime---down to $O(N^5)$ for sparse graphs.

%MIKE: May need several
More recent approaches for solving MinCost may yield improvements to the total run time. 
However, despite these developments, current state-of-the-art algorithms for solving MinCost still require significantly more computation than those for solving MaxCard.
Consequently, solving MinCost tends to be the computational bottleneck for statistical matching algorithms.

\begin{figure}
\begin{center}
\begin{minipage}{2 in}
\begin{tikzpicture}[scale=0.8] 
		%Sourse node
		\node (0) [circle, fill=green!30] at (-1, 3) {s};
		\node (5) [circle, fill=green!30] at (6, 3) {t};
		%Treated nodes
		\node (1)[circle, fill=red!30] at (1,4) {$T_1$};
		\node (2)[circle, fill=red!30] at (1,2) {$T_2$};

		\tikzstyle{arrow} = [thick,->,>=stealth]
		%Control nodes
		\node (3)[circle, fill=blue!30] at (4,4) {$C_1$};
		\node (4) [circle, fill=blue!30]at (4,2) {$C_2$};

\draw (1) edge[arrow, above][style = ultra thick, color = red][pos=0.2]node{$w_{11}$} (3)
(1) edge[arrow,above][pos=0.3]node{$w_{12}$} (4)
(2) edge[arrow,above][pos=0.1]node{$w_{21}$} (3)
(2) edge[arrow, below][style = ultra thick, color = red] [pos=0.2]node{$w_{22}$}(4)
(0) edge[arrow,above][style = ultra thick, color = red][pos=0.3]node{$1$} (1)
(0) edge[arrow,below][style = ultra thick, color = red][pos=0.3]node{$1$} (2)
(3) edge[arrow,above][style = ultra thick, color = red][pos=0.3]node{$1$} (5)
(4) edge[arrow,below][style = ultra thick, color = red][pos=0.3]node{$1$} (5)
;
\end{tikzpicture}\\ 
\centering
\textbf{(a)} 
\end{minipage}
\hspace{0.7 in}
\begin{minipage}{2 in}
\begin{tikzpicture}[scale=0.8]

			%Sourse node
		\node (0) [circle, fill=green!30] at (7, 3) {s};
		\node (5) [circle, fill=green!30] at (14, 3) {t};
		%Treated nodes
		%Treated nodes
		\node (1)[circle, fill=red!30]  at (9,4) {$T_1$};
		\node (2)[circle, fill=red!30]  at (9,2) {$T_2$};

		%Control nodes
		\node (3) [circle, fill=blue!30] at (12,4) {$C_1$};
		\node (4) [circle, fill=blue!30] at (12,2) {$C_2$};

\tikzstyle{arrow} = [thick,->,>=stealth]
	\draw 
(3) edge[arrow, above, bend right=10, color=blue] node{$-w_{11}$} (1)
(1) edge[arrow, below, bend right=10] node[pos=0.5]{$w_{11}$} (3)
(1) edge[arrow,below,color=blue]node[pos=0.1]{$w_{12}$} (4)
(2) edge[arrow,above,color=blue] node[pos=0.1]{$w_{21}$} (3)

(2) edge[arrow,above, bend left=10] node{$w_{22}$} (4)
(4) edge[arrow,below, bend left=10,color=blue] node{$-w_{22}$} (2)
(0) edge[arrow,below,bend right=10][pos=0.7]node{$1$} (1)
(1) edge[arrow, above, bend right=10] node{$-1$} (0)
(0) edge[arrow,above]node[pos=0.7]{$1$} (2)
(2) edge[arrow, below, bend left=20] node{$-1$} (0)
(3) edge[arrow,below,bend right=10][pos=0.2]node{$1$} (5)
(5) edge[arrow, above, bend right=10] node{$-1$} (3)
(4) edge[arrow,above,bend left=10][pos=0.3]node{$1$} (5)
(5) edge[arrow, below, bend left=20] node{$-1$} (4)
;
\end{tikzpicture}\\
\centering
\textbf{(b)} 
\end{minipage}

%\begin{center}

\begin{tikzpicture}[scale=0.8] 
		%Sourse node
		\node (0) [circle, fill=green!30] at (-1, 3) {s};
		\node (5) [circle, fill=green!30] at (6, 3) {t};
		%Treated nodes
		\node (1)[circle, fill=red!30] at (1,4) {$T_1$};
		\node (2)[circle, fill=red!30] at (1,2) {$T_2$};

		\tikzstyle{arrow} = [thick,->,>=stealth]
		%Control nodes
		\node (3)[circle, fill=blue!30] at (4,4) {$C_1$};
		\node (4) [circle, fill=blue!30]at (4,2) {$C_2$};

\draw (1) edge[arrow, above][pos=0.2]node{$w_{11}$} (3)
(1) edge[arrow,above][style = ultra thick, color = red][pos=0.3]node{$w_{12}$} (4)
(2) edge[arrow,above][style = ultra thick, color = red][pos=0.1]node{$w_{21}$} (3)
(2) edge[arrow, below] [pos=0.2]node{$w_{22}$}(4)
(0) edge[arrow,above][style = ultra thick, color = red][pos=0.3]node{$1$} (1)
(0) edge[arrow,below][style = ultra thick, color = red][pos=0.3]node{$1$} (2)
(3) edge[arrow,above][style = ultra thick, color = red][pos=0.3]node{$1$} (5)
(4) edge[arrow,below][style = ultra thick, color = red][pos=0.3]node{$1$} (5)
;
\end{tikzpicture}\\ 
\centering
\textbf{(c)}

	\caption{(a). The original flow network $G$ with initial flow $f$ with dissimilarities $w$. (b). The residual graph for (a) with augmenting path $p$ color in blue; consider the reverse-paths $c_1-T_1$ and $c_2-T_2$, selecting path $c_1-T_1-C_2-T_2-C_1$ with $-w_{11} + w_{12} - w_{22}+ w_{21} <0$ negative cost. (c). The flow in $G$ that results from augmenting along path $p$.}
	\label{fig4}
\end{center}
\end{figure}
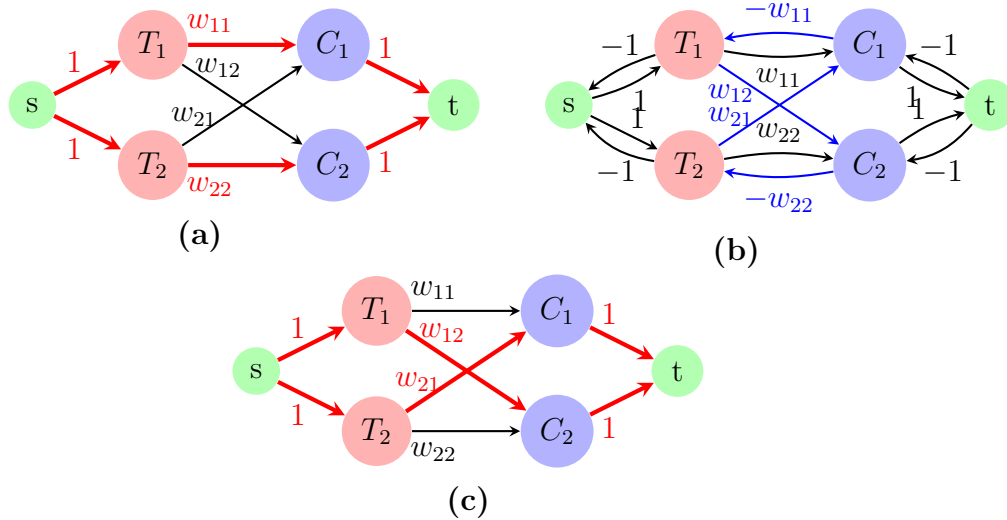

  \section{Scaling down data in statistical matching}
%This study first proposes a new algorithm to scale down data from an extensive data set. 
%First, it is essential to know why scaling down data is important in statistical matching problems

%  With the development of technology, the volume of big data has skyrocketed. 
%  The valuable insight of big data is essential for companies to improve operations and provide better service.
%  As in Figure 2:1(a), in a bipartite graph with $N_T$ number of treated units and $N_C$ number of control units, $N_T \times N_C$ number of edges can be drawn. Consider a data set with 1000 treatment units and 1000 control units, then $1000 \times 1000 $ edges can be drawn in the graph. In optimal bipartite matching, each edge is a binary decision variable. Considering the massive amount of binary decisions in a more extensive data set is challenging to complete. Let's understand some terminologies before answer why sparse graph is important in statistical matching.   

We have previously emphasized that potential gains in computational efficiency can be obtained by imposing sparsity in the matching graph $G$.   
Thus, as observational studies grow in size, the use of matching methods that perform a pre-processing step to sufficiently sparsify $G$ prior to matching seems critical.  
Ideally, the sparsification should be performed in a way to ensure that the matching solution on the sparse graph is similar to that on the original matching graph.  
While some matching methods that include this sparsification step have been developed, overall, there is still a substantial need for additional research in this area.  

% this is largely an underdeveloped area of research.  in need of significant additional work.  

% in dire need of additional research.

% have been developed that include this sparsification step, overall, the development of these types of algorithms are an area of  under sparsified matching graph $G$
% However, while some matching sparse matching methods are an underdeveloped area of research.

We now detail the logistics of matching on a sparse graph and give some examples of current techniques for imposing sparsity in the matching graph.
%matching methods that implement some level of sparsity in their solution.

% performrestricting attention to matching methods that include a sparsification step as well as sparsity of $G$ way to extend ideas from matching algorithms to larger studies.

% However, has not been widely studied, nor have algorithms commonly developed to explicitly solve statistical matching problems

% Overall, work to ensure while ensuring  our sparsity in the matching graph $G$

  \subsection{Matching on a Sparse Ggraph}

%FOR MIKE: TALK ABOUT EXSISTING METHODS: Bottleneck and K-nearest neighbor.

%FOR SANJEEWANI: Add in the relevant references to figures.
%As described in Section~\ref{sec:lap}, 
Matching graphs $G = (V,E)$ are often be expressed as an $N_T \times N_C$ cost matrix $W$---similar to the one constructed in Section~\ref{sec:lap}.
Cost matrices are easy to store as data and provide all the necessary information to perform a standard statistical matching algorithm.
%s require only a cost matrix as an input.

%FOR MIKE: TO EDIT
%Sanjeewani changed here
The cost matrix $W$ from a graph $G$ is constructed as follows.
If the edge $ij \in E$, then $W_{ij} = w_{ij}$.  
If $ij \notin E$, then $W_{ij} = \infty$ (or, in practice, is set to a number larger than any $w_{ij}$ for $ij \in E$).
%, as shown details in Figure ~\ref{fig7}, and ~\ref{fig6}.  
For this latter case, the large cost prevents algorithms from matching unit $i$ to $j$ instead of to $j'$ if $ij \notin E$ and $ij' \in E$ (provided both are possible).
It requires $O(N^2)$ memory to store a cost matrix.

\begin{figure}
    \centering

	\begin{tikzpicture}[scale=1.1]

		\tikzstyle{every node} = [circle, fill=red!30]
		%Treated nodes
		\node (1) at (1,4) {1};
		\node (2) at (1,3) {2};
		\node (3) at (1,2) {3};
		\node (4) at (1,1) {4};
		\node (5) at (1,0) {5};
		
		\tikzstyle{every node} = [circle, fill=blue!30]
		%Control nodes
		\node (6) at (4,4) {1};
		\node (7) at (4,3) {2};
		\node (8) at (4,2) {3};
		\node (9) at (4,1) {4};
		\node (10) at (4,0) {5};

            \draw (1) -- (6);
		\draw (1) -- (7);
			
		\draw (2) -- (7);
		\draw (2) -- (8);
		
		\draw (3) -- (8);
		\draw (3) -- (9);
	
		\draw (4) -- (9);
		\draw (4) -- (10);
  
		\draw (5) -- (6);
		\draw (5) -- (10);
	
		%[style = thick][->] [arrow] (\from) -- (\to);
	\end{tikzpicture}\\
	\caption{A sparse bipartite graph with five treated and control units} 
	\label{fig7}
 \end{figure}
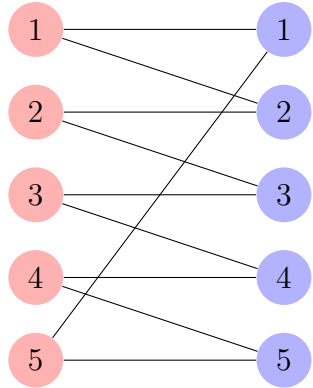

Note that, if $G$ is a complete bipartite graph, then $W$ will only have finite entries, and if $G$ is a \textit{dense} graph---that is, if the number of edges is proportional to $N^2$---then a significant proportion of entries will be finite.
However, if $G$ is sparse graph---that is, if the number of edges is proportional to $N$---then most of the entries of $W$ are infinite. 
That is, the bulk of the $O(N^2)$ memory required to store $W$ will be devoted to storing infinite values which will not be used when optimizing the matching algorithm.
Figure~\ref{fig7} provides an example of a sparse graph.

Instead, when matching problems are sparse, \textit{adjacency lists} tend to be the preferred object for storing the information in $G$.  
%MIKE: Would need a picture of an adjacency list with a cite.
For every node $i$, an adjacency list stores a vector $v_i$ containing all nodes $j$ which are incident to $i$.  
Edge costs can be stored, for example, within a second vector $v^w_i$, where the $\ell~\text{th}$ entry of $v^w_i$ is the cost between $i$ and the node in the $\ell~\text{th}$ entry of $v_i$.
% Edge costs can be stored using a second list, where the $\ell~\text{th}$ entry of the vector corresponding to unit $i$ entry of the through the addition of a separate corresponding list cont
If, on average, each node is incident to $k$ other nodes, then the memory requirement to store an adjacency list is $O(kN)$, which is significantly less than $O(N^2)$ for large-to-massive matching problems.
Figure~\ref{fig6} provides the representation of the graph in Figure~\ref{fig7} as both a cost matrix and as an adjacency list.

However, in smaller matching problems where memory is not an issue, cost matrices may be preferable to adjacency lists.
For example, when storing $G$ as a cost matrix $W$, determining whether an edge $ij \in E$ is performed by accessing $W_{ij}$ and checking whether it is finite---this operation requires $O(1)$ time.  
However, for an adjacency list, this operation requires inspecting all entries in $v_i$ to determine if $j \in v_i$, which requires $O(N)$ time. 
Additionally, matrix operations---for example, computing eigenvalues---may not be straight-forward using an adjacency list.

\subsection{Imposing Sparsity in a Matching Problem}

Currently, the most common way to impose sparsity in a matching problem is to prevent two units from being matched together if the corresponding cost of this match is prohibitively large.  
More rigorously, for a researcher-specified value of $\omega$, $i$ is only allowed to be matched to $j$ if $w_{ij} \leq \omega$.
% between units that have a cost of a researcher-specified value of $\omega$ or less.
In practice, this is known as imposing a \textit{bottleneck constraint}~\citep{hochbaum1986unified}
%MIKE: Check, is the minimax here on the maximum length
%, a minimax constraint, 
or a
%MIKE: Need another cite for caliper.
\textit{caliper}~\citep{rosenbaum1989optimal} on the matching.
%MIKE: CITE ROSENBAUM, GENERALIZED FULL MATCHING, SAM PIMENTEL, OTHER RECENT ROSENBAUM WORK, YOUR DISSERTATION
This type of sparsification can be performed fairly efficiently; a search through all possible matches requires $O(N^2)$ time.
Considerable recent work has devoted to implementing these types of constraints within matching problems.

% to prevent two matcj---this is known as imposing a \textit{bottleneck constraint}~\citep{hochbaum1986unified} or a 
% \textit{caliper}~\citep{rosenbaum1989optimal} on the edge costs.

Methods to find common support prior to matching may also be useful in reducing the total computational cost of matching.
%MIKE: CITE A FEW COMMON SUPPORT THINGS.  King and Zeng convex hull, Sharif's dissertation, Fogarty, maybe one article that provides a summary
Regions of common support are often much smaller than the entire population of units under study, and ensuring common support will often lead to a dramatic reduction in the number of control units (and possibly, the number of treated units) prior to matching.
However, most common support methods are not designed to impose sparsity---often, it is assumed that every treatment-control pair within the region of common support may be matched together---and additional steps are necessary to induce sparsity in the matching problem.

\begin{figure}
\centering
\begin{minipage}{2 in}
\begin{tikzpicture}[scale=1]
\centering

\matrix [matrix of nodes, left delimiter=(,right delimiter=) ]
{
        1&1&0&0&0\\
        0&1&1&0&0\\
        0&0&1&1&0\\
        0&0&0&1&1\\
        1&0&0&0&1\\
        };
\end{tikzpicture}
\vspace{-.01 in}
\hspace{-.7in}
%\centering
\textbf{}
\textbf{(a)}
\end{minipage}
\hspace{2.7in}

%MIKE: May not need b)
\begin{minipage}{2 in}
\begin{tikzpicture}[scale=1]
\centering

%MIKE: Use w not omega
\matrix [matrix of nodes, left delimiter=(,right delimiter=) ]
{
    $w_{11}$&$w_{12}$&$\infty$&$\infty$&$\infty$\\
    $\infty$&$w_{22}$&$w_{23}$&$\infty$&$\infty$\\
    $\infty$&$\infty$&$w_{33}$&$w_{34}$&$\infty$\\
    $\infty$&$\infty$&$\infty$&$w_{44}$&$w_{45}$\\
    $w_{51}$&$\infty$&$\infty$&$\infty$&$w_{55}$\\
};
\end{tikzpicture}

\centering
\vspace{-.03in}
\textbf{(b)}
\end{minipage}
\hspace{.7 in}

\centering
\begin{minipage}{2 in}
%\begin{tikzpicture}[scale=1]
\centering
\begin{tikzpicture}[every node/.style={minimum width=.9cm, minimum height=.7cm}]

\node[rectangle, draw, text = olive,fill=red!50] at (0,0) {$T_1$};

\node[rectangle, draw] at (1,0)    {$C_1$};
\node[rectangle, draw] at (1.9,0)    {$w_{11}$};
\node[rectangle, draw] at (3,0)    {$C_2$};
\node[rectangle, draw] at (3.9,0)    {$w_{12}$};

\node[rectangle, draw, text = olive,fill=red!50] at (0,-.8) {$T_2$};

\node[rectangle, draw] at (1,-.8)    {$C_2$};
\node[rectangle, draw] at (1.9,-.8)    {$w_{22}$};
\node[rectangle, draw] at (3,-.8)    {$C_3$};
\node[rectangle, draw] at (3.9,-.8)    {$w_{23}$};

\node[rectangle, draw, text = olive,fill=red!50] at (0,-1.6) {$T_3$};

\node[rectangle, draw] at (1,-1.6)    {$C_3$};
\node[rectangle, draw] at (1.9,-1.6)    {$w_{33}$};
\node[rectangle, draw] at (3,-1.6)    {$C_4$};
\node[rectangle, draw] at (3.9,-1.6)    {$w_{34}$};

\node[rectangle, draw, text = olive,fill=red!50] at (0,-2.4) {$T_4$};

\node[rectangle, draw] at (1,-2.4)    {$C_4$};
\node[rectangle, draw] at (1.9,-2.4)    {$w_{41}$};
\node[rectangle, draw] at (3,-2.4)    {$C_5$};
\node[rectangle, draw] at (3.9,-2.4)    {$w_{44}$};

\node[rectangle, draw, text = olive,fill=red!50] at (0,-3.2) {$T_5$};

\node[rectangle, draw] at (1,-3.2)    {$C_1$};
\node[rectangle, draw] at (1.9,-3.2)    {$w_{41}$};
\node[rectangle, draw] at (3,-3.2)    {$C_5$};
\node[rectangle, draw] at (3.9,-3.2)    {$w_{42}$};

\end{tikzpicture}
\centering
\textbf{(c)}
\end{minipage}
\hspace{0.7 in}
 \label{fig:three sin x}

\caption{(a). Adjacency matrix, (b). Cost matrix, and (c) Adjacency list for the sparse graph in Figure ~\ref{fig7} }
        \label{fig6}
\end{figure}
 
% In terms of the accessing time, an adjacency matrix is more efficient to check a given edge exists. However, in an adjacency list, it is exactly the opposite. All the list elements need to be checked to find any relationship with the list structure in a graph. Hence, an adjacency matrix is much more efficient when finding the connections in a graph in terms of the accessing time.
 
% The main advantage of adjacency lists over the matrix is space requirement. Since the adjacency matrix is a $N\times N$ matrix, it requires $O(N^2)$ space complexity. However, adjacency lists only store the node connected in a list or a vector. Thus, the adjacency list uses much lower memory than the matrices for graphs. Therefore, the adjacency list is much more efficient for storing the graph, especially sparse graphs, with fewer edges than nodes.

% %FOR MIKE: A name more like this.
 \section{Software for Statistical Matching}

There are a variety of software packages available for performing statistical matching without replacement, especially for the \texttt{R} programming language.  
Commonly used \texttt{R} packages include \texttt{Matching}~\citep{sekhon2008multivariate}, \texttt{MatchIt}~\citep{stuart2011matchit}, and \texttt{optmatch} \citep{hansen2007optmatch}. 
Additionally, a recently developed package,\\ \texttt{rcbalance}~\citep{pimentel2022package}, is explicitly designed to solve sparse matching problems, and allows users to input the statistical matching problem as an adjacency list.

%MIKE: START HERE
Under the hood, however, most of these packages tend to use the same handful of algorithms to solve optimal matching problems.  
Historically, the most commonly-used algorithm has been the \texttt{Relax-IV} algorithm~\citep{bertsekas1994relax}.  
This algorithm solves the matching problem using a coordinate ascent procedure on the dual of the assignment problem (see Section~\ref{sec:lap})~\citep{bertsekas1981new, bertsekas1988relaxation,bertsekas1988relax} where an initial solution is obtained via an auction algorithm~\citep{bertsekas1992auction}.
This algorithm is free to use for academic research purposes, but requires special permission for non-research or commercial uses.  
%Bertsekas and Tseng freely permit their software to be used for research purposes
Additionally, this algorithm has been largely unchanged since 1994.

The LEMON (Library for Efficient Modeling and Optimization in Networks) solver library has grown in recent popularity~\citep{dezsHo2011lemon}.
LEMON can solve a wide variety of optimization problems on graphs, and in particular, has four efficient implementations for solving instances of MinCost: cycle cancelling, network simplex, cost scaling, and capacity scaling. 
These implementations appear to perform competitively when compared to other implementations~\citep{kovacs2015minimum}.
Of particular note, LEMON is free and has a very permissive license that allows its use for both academic and commercial purposes.

Some statistical matching packages---for example,\\ \texttt{MatchIt} and \texttt{designmatch} \citep{zubizarreta2018designmatch}---allow for the use of the proprietary optimization libraries to solve the matching problem.
The most commonly used libraries include \texttt{Gurobi}~\citep{gurobi2021gurobi} and \texttt{CPLEX}~\citep{cplex2009v12}.
Like LEMON, these libraries are designed to efficiently solve a wide variety of linear and integer programming problems, not just those related to MinCost or LUAP.
However, these libraries are not free to use outside of academic purposes.
%solving of many different kinds of integer programming problems.

Finally, a potentially useful algorithm for solving statistical matching problems is the CS2 (cost-scaling 2) algorithm~\citep{goldberg1997efficient}, a type of push-relabel algorithm.  
% This is a cost-scaling push-relabel algorithm \citep{goldberg1997efficient}. The cost-scaling codes are substantially faster on the largest sparse network due to their better asymptotic behavior regarding the number of nodes. This variant of the cost-scaling algorithm performs path augmentations instead of local push operations, and relabeling is heavily used to find augmenting paths. 
% that the CS2 algorithm COST SCALING
Simulation studies have shown this algorithm to be one of the most efficient available at solving MinCost~\citep{kovacs2015minimum}.
CS2 appears to have been free to download and use for academic purposes, and some implementations of this algorithm can be found with a Google search.
% However, it does not appear to be 

%FOR MIKE: START HERE
\section{Statistical Matching on Massive Data Moving Forward}

Overall, there appears to be a need for further development and implementation of algorithms for solving optimal statistical matching problems.
Ideally, these algorithms should be tailored to take advantage of properties particular to the optimal matching problem---for example, if solving MinCost, that all edges have a capacity of 1.  
These algorithms may also benefit from smart choices of the dissimilarity measure.
For example, additional approaches may be available if the edge costs satisfy the triangle inequality~\citep{hochbaum1986unified}.

Finally, as statistical matching problems continue to grow in scale, the computational complexity of these problems will necessitate statistical matching techniques that impose sparsity on the matching problem.
Algorithms designed and implemented to exploit sparsity of the matching graph---for example, that in~\citet{axiotis2022faster}---seem ideal for these types of matching problems. 

\bibliographystyle{plainnat}
\bibliography{references}

\begin{thebibliography}{49}
\providecommand{\natexlab}[1]{#1}
\providecommand{\url}[1]{\texttt{#1}}
\expandafter\ifx\csname urlstyle\endcsname\relax
  \providecommand{\doi}[1]{doi: #1}\else
  \providecommand{\doi}{doi: \begingroup \urlstyle{rm}\Url}\fi

\bibitem[Agarwal and Sharathkumar(2014)]{agarwal2014approximation}
Pankaj~K Agarwal and R~Sharathkumar.
\newblock Approximation algorithms for bipartite matching with metric and
  geometric costs.
\newblock In \emph{Proceedings of the forty-sixth annual ACM symposium on
  Theory of computing}, pages 555--564, 2014.

\bibitem[Austin and Small(2014)]{austin2014use}
Peter~C Austin and Dylan~S Small.
\newblock The use of bootstrapping when using propensity-score matching without
  replacement: {A} simulation study.
\newblock \emph{Statistics in medicine}, 33\penalty0 (24):\penalty0 4306--4319,
  2014.

\bibitem[Axiotis et~al.(2022)Axiotis, Madry, and Vladu]{axiotis2022faster}
Kyriakos Axiotis, Aleksander Madry, and Adrian Vladu.
\newblock Faster sparse minimum cost flow by electrical flow localization.
\newblock In \emph{2021 IEEE 62nd Annual Symposium on Foundations of Computer
  Science (FOCS)}, pages 528--539. IEEE, 2022.

\bibitem[Bachem et~al.(1992)Bachem, Kern, Bachem, and Kern]{bachem1992linear}
Achim Bachem, Walter Kern, Achim Bachem, and Walter Kern.
\newblock \emph{Linear programming duality}.
\newblock Springer, 1992.

\bibitem[Bertsekas(1981)]{bertsekas1981new}
Dimitri~P Bertsekas.
\newblock A new algorithm for the assignment problem.
\newblock \emph{Mathematical Programming}, 21\penalty0 (1):\penalty0 152--171,
  1981.

\bibitem[Bertsekas and Tseng(1988{\natexlab{a}})]{bertsekas1988relax}
Dimitri~P Bertsekas and Paul Tseng.
\newblock The relax codes for linear minimum cost network flow problems.
\newblock \emph{Annals of Operations Research}, 13\penalty0 (1):\penalty0
  125--190, 1988{\natexlab{a}}.

\bibitem[Bertsekas and Tseng(1988{\natexlab{b}})]{bertsekas1988relaxation}
Dimitri~P Bertsekas and Paul Tseng.
\newblock Relaxation methods for minimum cost ordinary and generalized network
  flow problems.
\newblock \emph{Operations research}, 36\penalty0 (1):\penalty0 93--114,
  1988{\natexlab{b}}.

\bibitem[Bertsekas et~al.(1994)Bertsekas, Tseng, et~al.]{bertsekas1994relax}
Dimitri~P Bertsekas, Paul Tseng, et~al.
\newblock Relax-iv: A faster version of the relax code for solving minimum cost
  flow problems.
\newblock 1994.

\bibitem[Bertsekas et~al.(1992)]{bertsekas1992auction}
Dimitri~P Bertsekas et~al.
\newblock An auction/sequential shortest path algorithm for the minimum cost
  network flow problem.
\newblock \emph{Report P-2146}, 1992.

\bibitem[Bijsterbosch and Volgenant(2010)]{bijsterbosch2010solving}
J~Bijsterbosch and A~Volgenant.
\newblock Solving the rectangular assignment problem and applications.
\newblock \emph{Annals of Operations Research}, 181\penalty0 (1):\penalty0
  443--462, 2010.

\bibitem[Bottigliengo et~al.(2021)Bottigliengo, Baldi, Lanera, Lorenzoni,
  Bejko, Bottio, Tarzia, Carrozzini, Gerosa, Berchialla,
  et~al.]{bottigliengo2021oversampling}
Daniele Bottigliengo, Ileana Baldi, Corrado Lanera, Giulia Lorenzoni, Jonida
  Bejko, Tomaso Bottio, Vincenzo Tarzia, Massimiliano Carrozzini, Gino Gerosa,
  Paola Berchialla, et~al.
\newblock Oversampling and replacement strategies in propensity score matching:
  a critical review focused on small sample size in clinical settings.
\newblock \emph{BMC medical research methodology}, 21\penalty0 (1):\penalty0
  1--16, 2021.

\bibitem[Burkard et~al.(2012)Burkard, Dell'Amico, and
  Martello]{burkard2012assignment}
Rainer Burkard, Mauro Dell'Amico, and Silvano Martello.
\newblock \emph{Assignment problems: revised reprint}.
\newblock SIAM, 2012.

\bibitem[Cormen et~al.(2022)Cormen, Leiserson, Rivest, and
  Stein]{cormen2022introduction}
Thomas~H Cormen, Charles~E Leiserson, Ronald~L Rivest, and Clifford Stein.
\newblock \emph{Introduction to algorithms}.
\newblock MIT press, 2022.

\bibitem[{CPLEX}(2009)]{cplex2009v12}
{IBM}~{ILOG} {CPLEX}.
\newblock V12. 1: User’s manual for {CPLEX}.
\newblock \emph{International Business Machines Corporation}, 46\penalty0
  (53):\penalty0 157, 2009.

\bibitem[Dantzig(1990)]{dantzig1990origins}
George~B Dantzig.
\newblock Origins of the simplex method.
\newblock In \emph{A history of scientific computing}, pages 141--151. 1990.

\bibitem[Dasgupta et~al.(2008)Dasgupta, Papadimitriou, and
  Vazirani]{dasgupta2008algorithms}
Sanjoy Dasgupta, Christos~H Papadimitriou, and Umesh~Virkumar Vazirani.
\newblock \emph{Algorithms}.
\newblock McGraw-Hill Higher Education New York, 2008.

\bibitem[Dehejia and Wahba(2002)]{dehejia2002propensity}
Rajeev~H Dehejia and Sadek Wahba.
\newblock Propensity score-matching methods for nonexperimental causal studies.
\newblock \emph{Review of Economics and statistics}, 84\penalty0 (1):\penalty0
  151--161, 2002.

\bibitem[Dezs{\H{o}} et~al.(2011)Dezs{\H{o}}, J{\"u}ttner, and
  Kov{\'a}cs]{dezsHo2011lemon}
Bal{\'a}zs Dezs{\H{o}}, Alp{\'a}r J{\"u}ttner, and P{\'e}ter Kov{\'a}cs.
\newblock Lemon--an open source c++ graph template library.
\newblock \emph{Electronic notes in theoretical computer science}, 264\penalty0
  (5):\penalty0 23--45, 2011.

\bibitem[Diamond and Sekhon(2013)]{diamond2013genetic}
Alexis Diamond and Jasjeet~S Sekhon.
\newblock Genetic matching for estimating causal effects: A general
  multivariate matching method for achieving balance in observational studies.
\newblock \emph{Review of Economics and Statistics}, 95\penalty0 (3):\penalty0
  932--945, 2013.

\bibitem[Dutta and Pal(2015)]{dutta2015note}
Jayanta Dutta and SC~Pal.
\newblock A note on hungarian method for solving assignment problem.
\newblock \emph{Journal of Information and Optimization Sciences}, 36\penalty0
  (5):\penalty0 451--459, 2015.

\bibitem[Edmonds(1965{\natexlab{a}})]{edmonds1965maximum}
Jack Edmonds.
\newblock Maximum matching and a polyhedron with 0, 1-vertices.
\newblock \emph{Journal of research of the National Bureau of Standards B},
  69\penalty0 (125-130):\penalty0 55--56, 1965{\natexlab{a}}.

\bibitem[Edmonds(1965{\natexlab{b}})]{edmonds1965paths}
Jack Edmonds.
\newblock Paths, trees, and flowers.
\newblock \emph{Canadian Journal of mathematics}, 17:\penalty0 449--467,
  1965{\natexlab{b}}.

\bibitem[Fang and Gong(2017)]{fang2017extended}
Donghui Fang and Xin Gong.
\newblock Extended farkas lemma and strong duality for composite optimization
  problems with dc functions.
\newblock \emph{Optimization}, 66\penalty0 (2):\penalty0 179--196, 2017.

\bibitem[Ford and Fulkerson(1957)]{ford1957simple}
Lester~Randolph Ford and Delbert~R Fulkerson.
\newblock A simple algorithm for finding maximal network flows and an
  application to the hitchcock problem.
\newblock \emph{Canadian journal of Mathematics}, 9:\penalty0 210--218, 1957.

\bibitem[Gliklich et~al.(2019)Gliklich, Leavy, and Dreyer]{gliklich2019tools}
Richard~E Gliklich, Michelle~B Leavy, and Nancy~A Dreyer.
\newblock Tools and technologies for registry interoperability, registries for
  evaluating patient outcomes: A user’s guide, addendum 2 [internet].
\newblock 2019.

\bibitem[Goldberg(1997)]{goldberg1997efficient}
Andrew~V Goldberg.
\newblock An efficient implementation of a scaling minimum-cost flow algorithm.
\newblock \emph{Journal of algorithms}, 22\penalty0 (1):\penalty0 1--29, 1997.

\bibitem[Goldberg and Tarjan(1989)]{goldberg1989finding}
Andrew~V Goldberg and Robert~E Tarjan.
\newblock Finding minimum-cost circulations by canceling negative cycles.
\newblock \emph{Journal of the ACM (JACM)}, 36\penalty0 (4):\penalty0 873--886,
  1989.

\bibitem[Gurobi~Optimization(2021)]{gurobi2021gurobi}
LLC Gurobi~Optimization.
\newblock Gurobi optimizer reference manual, 2021.

\bibitem[Hansen(2007)]{hansen2007optmatch}
Ben~B Hansen.
\newblock Optmatch: Flexible, optimal matching for observational studies.
\newblock \emph{New Functions for Multivariate Analysis}, 7\penalty0
  (2):\penalty0 18--24, 2007.

\bibitem[Hochbaum and Shmoys(1986)]{hochbaum1986unified}
Dorit~S Hochbaum and David~B Shmoys.
\newblock A unified approach to approximation algorithms for bottleneck
  problems.
\newblock \emph{Journal of the ACM (JACM)}, 33\penalty0 (3):\penalty0 533--550,
  1986.

\bibitem[Imbens and Rubin(2015)]{imbens2015causal}
Guido~W Imbens and Donald~B Rubin.
\newblock \emph{Causal inference in statistics, social, and biomedical
  sciences}.
\newblock Cambridge University Press, 2015.

\bibitem[Khachiyan and Porkolab(2000)]{khachiyan2000integer}
Leonid Khachiyan and Lorant Porkolab.
\newblock Integer optimization on convex semialgebraic sets.
\newblock \emph{Discrete \& Computational Geometry}, 23\penalty0 (2):\penalty0
  207--224, 2000.

\bibitem[Klein(1967)]{klein1967primal}
Morton Klein.
\newblock A primal method for minimal cost flows with applications to the
  assignment and transportation problems.
\newblock \emph{Management Science}, 14\penalty0 (3):\penalty0 205--220, 1967.

\bibitem[K{\H{o}}nig(1931)]{kHonig1931grafok}
D{\'e}nes K{\H{o}}nig.
\newblock Gr{\'a}fok {\'e}s m{\'a}trixok.
\newblock \emph{Matematikai {\'e}s Fizikai Lapok}, 38:\penalty0 116--119, 1931.

\bibitem[Kov{\'a}cs(2015)]{kovacs2015minimum}
P{\'e}ter Kov{\'a}cs.
\newblock Minimum-cost flow algorithms: an experimental evaluation.
\newblock \emph{Optimization Methods and Software}, 30\penalty0 (1):\penalty0
  94--127, 2015.

\bibitem[Kuhn(1955)]{kuhn1955hungarian}
Harold~W Kuhn.
\newblock The hungarian method for the assignment problem.
\newblock \emph{Naval research logistics quarterly}, 2\penalty0 (1-2):\penalty0
  83--97, 1955.

\bibitem[Munkres(1957)]{munkres1957algorithms}
James Munkres.
\newblock Algorithms for the assignment and transportation problems.
\newblock \emph{Journal of the society for industrial and applied mathematics},
  5\penalty0 (1):\penalty0 32--38, 1957.

\bibitem[Nelder and Mead(1965)]{nelder1965simplex}
John~A Nelder and Roger Mead.
\newblock A simplex method for function minimization.
\newblock \emph{The computer journal}, 7\penalty0 (4):\penalty0 308--313, 1965.

\bibitem[Pimentel et~al.(2022)Pimentel, Pimentel, et~al.]{pimentel2022package}
Samuel~D Pimentel, Maintainer Samuel~D Pimentel, et~al.
\newblock rcbalance: Large, sparse optimal matching with refined covariate
  balance.
\newblock 2022.

\bibitem[Radzik and Goldberg(1994)]{radzik1994tight}
Tomasz Radzik and Andrew~V Goldberg.
\newblock Tight bounds on the number of minimum-mean cycle cancellations and
  related results.
\newblock \emph{Algorithmica}, 11\penalty0 (3):\penalty0 226--242, 1994.

\bibitem[Reingold and Tarjan(1981)]{reingold1981greedy}
Edward~M Reingold and Robert~E Tarjan.
\newblock On a greedy heuristic for complete matching.
\newblock \emph{SIAM Journal on Computing}, 10\penalty0 (4):\penalty0 676--681,
  1981.

\bibitem[Rosenbaum(1989)]{rosenbaum1989optimal}
Paul~R Rosenbaum.
\newblock Optimal matching for observational studies.
\newblock \emph{Journal of the American Statistical Association}, 84\penalty0
  (408):\penalty0 1024--1032, 1989.

\bibitem[Rosenbaum et~al.(2010)Rosenbaum, Rosenbaum, and
  Briskman]{rosenbaum2010design}
Paul~R Rosenbaum, P~Rosenbaum, and Briskman.
\newblock \emph{Design of observational studies}, volume~10.
\newblock Springer, 2010.

\bibitem[S{\"a}vje et~al.(2021)S{\"a}vje, Higgins, and
  Sekhon]{savje2021generalized}
Fredrik S{\"a}vje, Michael~J Higgins, and Jasjeet~S Sekhon.
\newblock Generalized full matching.
\newblock \emph{Political Analysis}, 29\penalty0 (4):\penalty0 423--447, 2021.

\bibitem[Sekhon(2008)]{sekhon2008multivariate}
Jasjeet~S Sekhon.
\newblock Multivariate and propensity score matching software with automated
  balance optimization: the matching package for r.
\newblock \emph{Journal of Statistical Software, Forthcoming}, 2008.

\bibitem[Stuart et~al.(2011)Stuart, King, Imai, and Ho]{stuart2011matchit}
Elizabeth~A Stuart, Gary King, Kosuke Imai, and Daniel Ho.
\newblock Matchit: nonparametric preprocessing for parametric causal inference.
\newblock \emph{Journal of statistical software}, 2011.

\bibitem[Zubizarreta(2012)]{zubizarreta2012using}
Jos{\'e}~R Zubizarreta.
\newblock Using mixed integer programming for matching in an observational
  study of kidney failure after surgery.
\newblock \emph{Journal of the American Statistical Association}, 107\penalty0
  (500):\penalty0 1360--1371, 2012.

\bibitem[Zubizarreta et~al.(2014)Zubizarreta, Paredes, Rosenbaum,
  et~al.]{zubizarreta2014matching}
Jos{\'e}~R Zubizarreta, Ricardo~D Paredes, Paul~R Rosenbaum, et~al.
\newblock Matching for balance, pairing for heterogeneity in an observational
  study of the effectiveness of for-profit and not-for-profit high schools in
  chile.
\newblock \emph{Annals of Applied Statistics}, 8\penalty0 (1):\penalty0
  204--231, 2014.

\bibitem[Zubizarreta et~al.(2018)Zubizarreta, Kilcioglu, and
  Vielma]{zubizarreta2018designmatch}
JR~Zubizarreta, C~Kilcioglu, and JP~Vielma.
\newblock designmatch: Matched samples that are balanced and representative by
  design.
\newblock \emph{R package version 0.3}, 1, 2018.

\end{thebibliography}

\end{document}